\documentclass{aa}  

\usepackage{graphicx}
\usepackage{txfonts}
\usepackage{amsmath}
\usepackage{hyperref}
\newcommand{\code}[1]{\texttt{#1}\xspace}
\begin{document}

   \title{The $R$-Process Alliance: Analysis of Limited-$r$ Stars}

   \author{T. Xylakis-Dornbusch\inst{1}
          \and
          T. T. Hansen\inst{2}
          \and
          T. C. Beers\inst{3,4}
          \and
          N. Christlieb\inst{1}
          \and
          R. Ezzeddine\inst{5}
          \and
          A. Frebel\inst{6,4}
          \and
          E. Holmbeck\inst{7}
          \and
          V. M. Placco\inst{8}
          \and
          I. U. Roederer\inst{9,4}
          \and
          C. M. Sakari\inst{10}
          \and 
          C. Sneden\inst{11}
          }

   \institute{Zentrum f\"ur Astronomie der Universit\"at Heidelberg, Landessternwarte, K\"onigstuhl 12, 69117 Heidelberg, Germany\\
              \email{txylaki@lsw.uni-heidelberg.de}
         \and
         Department of Astronomy, Stockholm University, AlbaNova
University Center, SE-106 91 Stockholm, Sweden
             \and
             Department of Physics, University of Notre Dame, Notre Dame, IN 46556, USA
            \and
            JINA Center for the Evolution of the Elements, USA
            \and
            Department of Astronomy, University of Florida, 211 Bryant Space Science Center, Gainesville, FL 32601, USA
            \and
            Department of Physics and Kavli Institute for Astrophysics and
Space Research, Massachusetts Institute of Technology, Cambridge, MA 02139, USA
\and
The Observatories of the Carnegie Institution for Science, 813 Santa Barbara St, Pasadena, CA 91101, USA
\and
NSF’s NOIRLab, Tucson, AZ 85719, USA
\and
Department of Physics, North Carolina State University, Raleigh, NC 27695, USA
\and
Department of Physics \& Astronomy, San Francisco State University, San Francisco CA 94132, USA
\and
 Department of Astronomy and McDonald Observatory, The University of Texas, Austin, TX 78712, USA
             }
\date{}

 
  \abstract
   {In recent years, the $R$-Process Alliance (RPA) has conducted a successful search for stars enhanced in elements produced by the rapid neutron-capture ($r$-)process. In particular, the RPA has uncovered a number of stars strongly enriched in light $r$-process elements, such as Sr, Y and Zr, the so-called limited-$r$ stars, in order to investigate the astrophysical production site(s) of these elements.}
   {With this paper, we aim to investigate the possible formation sites for light neutron-capture elements, by deriving detailed abundances for neutron-capture elements from high-resolution, high signal-to-noise spectra of three limited-$r$ stars.}
   {We conducted a 1D local thermodynamic equilibrium (LTE) spectroscopic abundance analysis of three stars, as well as a kinematic analysis. Further, we calculated the lanthanide mass fraction ($X_{La}$) of our stars and of limited-$r$ stars from the literature.}
   {We found that the neutron-capture element abundance pattern of limited-$r$ stars behaves differently depending on their $\mathrm{[Ba/Eu]}$ ratios, and suggest that this should be taken into account in future investigations of their abundances. Furthermore, we found that the $X_{La}$ of limited-$r$ stars is lower than that of the kilonova AT2017gfo. The latter seems to be in the transition zone between limited-$r$ $X_{La}$ and that of $r$-I, $r$-II stars. Finally, we found that, unlike $r$-I and $r$-II stars, the current sample of limited-$r$ stars are largely born in the Galaxy rather than being accreted.} 
   {}


   \maketitle
%

\section{Introduction}\label{sec:intro}
Since the pioneering work of \cite{1957RvMP...29..547B} and \cite{Cameron_1957}, astronomers have known that elements beyond the iron peak are formed via the slow- and rapid-neutron-capture processes ($s$-process and $r$-process, respectively). However, the astrophysical site for the $r$-process is still highly debated. It has been hypothesized that two types of $r$-processes or two distinct sites may exist, differing by the available neutron flux; a main $r$-process in which all elements up to uranium can be produced and a neutron starved, so-called limited $r$-process, in which only the lighter elements can be formed (up to $\sim$Ba) \citep{frebel2018}. This limited $r$-process, also referred to as the weak $r$-process \citep{2012A&A...545A..31H} or the light element primary process (LEPP) \citep{Travaglio_2004}, was introduced in order to explain the observed abundance distribution of light $r$-process elements ($32<Z<56$) in metal-poor stars, which differs from the behaviour of the heavier elements. 

The $r$-process-enhanced (RPE) stars are divided into two sub-categories, namely $r$-I and $r$-II stars, for which $+0.3<\mathrm{[Eu/Fe]}\leq+0.7$ and $\mathrm{[Ba/Eu]}<0.0$, and $\mathrm{[Eu/Fe]}>+0.7$ and $\mathrm{[Ba/Eu]}<0.0$, respectively  \citep{2004A&A...428.1027C,beers2005,holmbeck2020}. Multiple studies have found that, for the $r$-I and $r$-II stars, a remarkable match is seen between the abundances of the old, metal-poor stars and the Sun for elements between the second and third $r$-process peaks ($55<Z<73$) \citep{sneden2008,Cowan_2021}. However, this universality does not extend to the lighter elements, where a larger scatter is seen. In particular, some stars are found that display an enhancement in the light $r$-process elements compared to the heavy ones, which is evident when scaled to the Solar-System $r$-process abundance pattern. These stars are characterized by the following abundance ratios: $\mathrm{[Eu/Fe]}<+0.3$, $\mathrm{[Sr/Ba]}>+0.5$ and $\mathrm{[Sr/Eu]}>0.0$, and are labelled limited-$r$ ($r_{lim}$) stars \citep{frebel2018}. 

The first star to be discovered displaying this type of neutron-capture element abundance pattern was HD~122563 \citep{1983ApJ...267..757S,2006ApJ...643.1180H,2007ApJ...666.1189H}. This star was found to exhibit a neutron-capture element abundance pattern which gradually decreases with growing atomic number. This was unlike any abundance pattern seen previously, and dissimilar to the pattern seen in $r$-I and $r$-II stars. The main question astronomers have tried to answer since the discovery of this difference in $r$-process stars is: Are the limited and main $r$-process components results of different events, or are they the product of the same event where different initial conditions or locations dictate the extent of the range of elements produced? 

The production sites of the $r$-process elements are still speculative, with the exception of neutron star mergers (NSMs), which were confirmed as such after the observation of the kilonova (KN) AT2017gfo, which was the electromagnetic counterpart \citep{2017Sci...358.1556C} of the gravitational event GW170817 \citep{2017PhRvL.119p1101A,2017ApJ...848L..12A}. Other candidate sites are collapsars \citep{Siegel_2019,Brauer_2021}, which are fast-rotating massive stars that end their lives as supernovae (SNe), magneto-rotational (MR) core-collapse supernovae (CCSN) \citep{2012ApJ...750L..22W}, and quark deconfinement SNe \citep{Fischer_2018,Fischer_2020}. The two former sites can produce, theoretically, both the main and limited components of the $r$-process, whereas the latter is a limited $r$-process candidate. Specifically, \cite{2017ApJ...836L..21N} found that MR-SNe that are driven by magneto-rotational instability can produce a variety of $r$-process patterns that range from the limited-$r$ to the Solar $r$-process pattern, when neutrino heating and magnetic fields are similar.

To investigate the limited-$r$ neutron-capture element abundance signature, and thereby constrain the possible production sites for these elements, the $R$-Process Alliance (RPA) has included these stars in their search in addition to identifying highly $r$-process-enhanced stars. Following \cite{frebel2018}, the RPA selected stars with $\mathrm{[Eu/Fe]}<+0.3$, $\mathrm{[Sr/Ba]}>+0.5$ and $\mathrm{[Sr/Eu]}>0.0$ as $r_{lim}$; in the first four data releases the RPA discovered 42 stars new $r_{lim}$ stars \citep{hansen2018,2018ApJ...868..110S,2020ApJ...898..150E,holmbeck2020}. This paper reports on the first detailed analysis of three of these $r_{lim}$ stars. The paper is organized as follows. In Section \ref{observations}, we describe the observations of the stars and in Section \ref{params_analysis}, we report on the stellar parameters and elemental abundances determination. The results are presented in Section \ref{results}. In Section \ref{discussion}, we discuss the possible birthplace of the $r_{lim}$ stars, and whether or not NSMs could be the production site for the elements observed in the atmospheres of these stars. 

\begin{table*}[ht]
\caption{Basic data for the sample stars.}
\label{tab:photo}
    \centering
    \resizebox{\textwidth}{!}{%
    \begin{tabular}{lllrrrrrrrrl}
    \hline\hline
    Stellar ID & RA & DEC & $B$ & $V$ & $J$ & $H$ & $K$ & $E(B-V)$ & BC$_v$ & $\varpi$ & $D$\\
               &    &     & mag & mag & mag & mag & mag & mag      &mag     & mas     & pc\\
   \hline
  2MASSJ00385967+2725516 & 00:39:00.2 & +27:25:33.9 & 12.18 & 11.44 &  9.87 &  9.40 &  9.35 & 0.04 & $-$0.37 & 0.78$\pm$0.02 & 1236$^{+42}_{-32}$\\
  2MASSJ20313531-3127319 & 20:31:35.0 & $-$31:27:24.3 & 14.36 & 13.57 & 11.94 & 11.47 & 11.37 & 0.08 & $-$0.49 & 0.38$\pm$0.02 & 2365$^{+76}_{-114}$\\
  2MASSJ21402305-1227035 & 21:40:23.3 & $-$12:26:59.8 & 11.94 & 11.04 &  9.23 &  8.76 &  8.62 & 0.05 & $-$0.51 & 0.34$\pm$0.03 & 2669$^{+197}_{-185}$\\  
\hline
    \end{tabular}}
    \tablebib{References: $B$ and $V$ magnitudes are from APASS \citep{2018AAS...23222306H} and 2MASS $JHK$ magnitudes were taken from \citet{cutri2003}. $E(B-V)$ is calculated using the dust maps from \citet{schlafly2011}, the bolometric corrections, BC$_v$, are based on \citet{casagrande2014}, the distances, $D$, are from \citet{bailerjones2018}, and the parallaxes, $\varpi$, from \cite{2023A&A...674A...1G}.}
   \end{table*}

\section{Observations\label{observations}}
\begin{figure*}
    \centering
\includegraphics[scale=0.45]{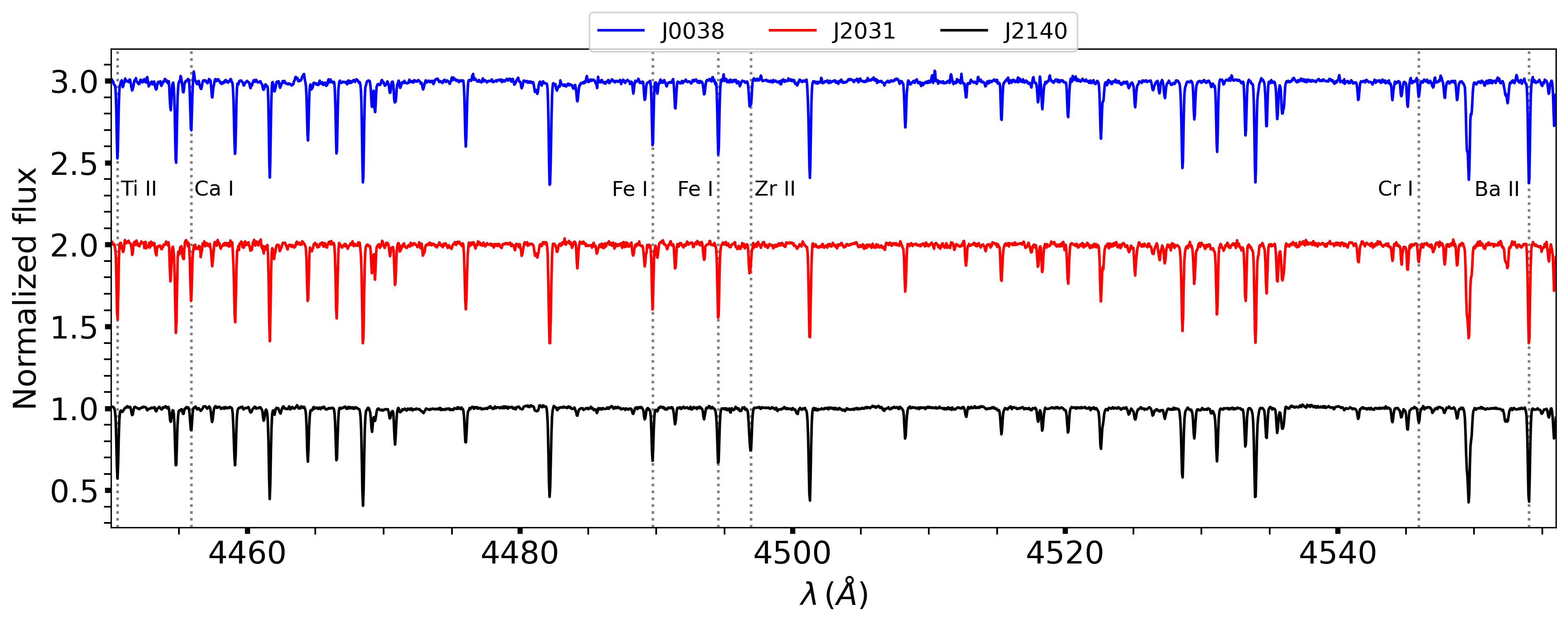}
    \caption{A snippet of the spectra for our three stars. Stars are offset in the direction of the y-axis to avoid overlap.}
    \label{fig:snippet} 
\end{figure*}

Our sample stars listed in Table \ref{tab:photo} were observed as part of the RPA survey for RPE stars. First, snapshot spectra were obtained ($R \sim 30,000$ and signal-to-noise ratio (SNR) $\sim30$ at 4100 \AA, see \citealt{hansen2018} for details) and analysed. Analysis of the snapshot spectra of J20313531-3127319 (J2031) and J21402305-1227035 (J2140) were published in \cite{hansen2018} and \cite{holmbeck2020}, respectively, while this paper presents the first analysis of J00385967+2725516 (J0038). Following analysis of the snapshot spectra, the three stars were selected as portrait candidates. Higher-resolution, higher SNR portrait spectra of J2031 and J2140 were obtained with the MIKE spectrograph \citep{2003SPIE.4841.1694B} on the Magellan/Clay telescope at the Las Campanas Observatory in Chile in April 2019, while the portrait spectrum of J0038 was obtained with the TS23 echelle spectrograph \citep{1995PASP..107..251T} on the Harlan J. Smith 107-in (2.7 m) telescope at McDonald Observatory in August 2020. The MIKE spectra cover a wavelength range of 3350 \AA\, to 5000 \AA\, in the blue and 4900 \AA\, to 9500 \AA\, in the red. The observations were obtained with a 0.7 x 5.0" slit and 2x2 binning, yielding a resolving power of $R\sim$37000 and  $R\sim$30000 in the blue and red, respectively. The McDonald spectra cover a wavelength range from 3400 \AA\, to 10900 \AA, and were obtained with the 1.8" slit and 1x1 binning, yielding a resolving power of $R\sim$35000. A snippet of all three spectra around 4500 \AA\, is shown in Figure \ref{fig:snippet}. As depicted, the quality of our spectra is ideal for accurate elemental abundances determination. The MIKE data was reduced with the CarPy MIKE pipeline \citep{kelson2000,kelson2003}, and the McDonald data was reduced using standard IRAF packages \citep{tody1986,tody1993}, including correction for bias, flatfield, and scattered light. Multiple spectra of the same star from different nights were subsequently co-added.  Table \ref{tab:photo} lists the target stellar identification (Stellar ID), right ascension (RA), and declination (DEC), while Table \ref{table:observs} lists the Heliocentric Julian Date (HJD), exposure times, SNR per pixel, and heliocentric radial velocities for the spectra. Heliocentric radial velocities of the stars were determined via cross-correlation of the object spectra with spectra of the standard star HD~122563 ($V_{helio}$=$-26.13$~$\mathrm{km\,s^{-1}}$ \citealt{gaia2018}) obtained with the same instruments. Thirty-five orders were used for the cross-correlation of the McDonald spectrum and fifty-five orders in the MIKE spectra, resulting in the mean radial velocities and standard deviations listed in Table \ref{table:observs}. All three stars have radial velocities reported in the literature. For J2031 and J2140, our velocities are consistent with previous measurements (J2031: $-$221.0~$\mathrm{km\,s^{-1}}$; \citealt{kunder2017}, $-$222.5~$\mathrm{km\,s^{-1}}$; \citealt{hansen2018}, and $-$221.1~$\mathrm{km\,s^{-1}}$ \citealt{steinmetz2020}. J2140: $-$133.0~$\mathrm{km\,s^{-1}}$; \citealt{beers2017}, and $-$130.4~$\mathrm{km\,s^{-1}}$; \citealt{gaia2018}). However, for J0038 a velocity of $-$97.56~$\mathrm{km\,s^{-1}}$ was reported by \cite{gaia2018}, which is $\sim20$ $\mathrm{km\,s^{-1}}$ less blue-shifted than what we find it to be, suggesting this star is part of a binary system. This assumption is supported by the fact that J0038 is included in the $Gaia$ DR3 non-single star (NSS) \code{nss\_twobody\_orbit} table \citep{Halbwachs2023}.

\begin{table}
        \caption{Observing log.}
        \label{table:observs}
    \centering
    \begin{tabular} {l c r r c}
    \hline\hline
    
    Object & HJD & Exposure time & SNR & $V_{helio}$ \\
     &  & (sec) & @4500 \AA & ($\mathrm{km\,s^{-1}}$)\\
    \hline
    J0038 & 2459087 & 5x1800 & 85$^*$ &$-$117.5$\pm$0.2\\
          & 2459088 & 5x1800 &        &$-$116.0$\pm$0.2\\
    J2031 & 2458600 & 4x900  & 87     &$-$220.6$\pm$0.3\\
    J2140 & 2458601 & 3x900  & 197    &$-$129.5$\pm$0.4\\
        \hline
        \end{tabular}%
    \tablefoot{$^*$ SNR of co-added spectra.}
\end{table}

\section{Stellar Parameters and Abundance Analysis}\label{params_analysis}
We used the software \code{smhr}\footnote{\href{https://github.com/andycasey/smhr}{https://github.com/andycasey/smhr}} \citep{2014Casey,2020AJ....160..181J} to normalise and then merge the orders of the echelle spectra. Then, we used it to fit Gaussians to measure the equivalent widths (EW) of spectral absorption lines.  Lastly, with \code{smhr} we derived the respective abundances from the curve of growth or from spectral synthesis via the 1D Local Thermodynamic Equilibrium (LTE) radiative transfer code \code{MOOG} (\cite{1973,sobeck2011}, 2017 version \footnote{\href{https://github.com/alexji/moog17scat}{https://github.com/alexji/moog17scat}}).

Parameters for the stars; effective temperature ($T_{\rm eff}$), surface gravity ($\log g$), metallicity ($\mathrm{[Fe/H]}$), and microturbulence ($\xi$), were determined following the procedure outlined in \citet{Roederer_2018b}. The $T_{\rm eff}$ for the stars were determined photometrically, that is, from the colours listed in Table \ref{tab:photo} using the colour-temperature relations of \cite{2010A&A...512A..54C}. Those were de-reddened using the \cite{schlafly2011} dust maps and extinction coefficients from \cite{2004AJ....128.2144M}. Further, the $\log g$ was calculated using the following fundamental relation: 
\begin{multline*}
    \mathrm{\log (g/g_{\odot})}=\log(M/\mathrm{M_{\odot}})-4\log\mathrm{(T_{eff}/T_{eff,\odot})}+0.4(M_{bol}-\mathrm{M_{bol,\odot}}) \\
    \mathrm{where\,} M_{\mathrm{bol}}=\mathrm{BC_v+V}+5\log\varpi+5-3.1E(B-V),
\end{multline*} 
using $\mathrm{M_{bol,\odot}} = 4.75$, $\log T_{\rm eff,\odot} = 3.7617$, and $\log g_\odot = 4.438$, and the parallaxes, $\varpi$, listed in Table \ref{tab:photo}.
Finally, EW measurements of \ion{Fe}{i} and \ion{Fe}{ii} lines were used to determine the metallicities and $\xi$. We adopt the $\mathrm{[FeI/H]}$ abundance as the model metallicity, and $\xi$ is the value ensuring that the \ion{Fe}{i} abundances are independent of their respective reduced equivalent widths. For all three stars the $\mathrm{[FeI/H]}$ and $\mathrm{[FeII/H]}$ abundances agree to within 0.03~dex. The final stellar parameters for the stars, along with associated uncertainties, are listed in Table \ref{table:astro_param}. In Table \ref{table:astro_param}, we list the combined systematic parameter uncertainties (see \citealt{Roederer_2018b} for details) and statistical uncertainties arising from the scatter in individual Fe-line abundances.

\begin{table}
        \caption{Stellar parameters of target stars.}
        \label{table:astro_param}
    \centering
    \resizebox{\columnwidth}{!}{%
    \begin{tabular} {c c c c c}
    \hline\hline
    
    Object & $\mathrm{T_{eff}}$ & $\mathrm{\log g}$ & $\mathrm{[Fe/H]}$ & $\xi$\\
     & (K) &  &  & ($\mathrm{km\,s^{-1}}$)\\
    \hline
    J0038 & 5203$\pm79$ & 2.45$\pm0.09$ & $-2.39\pm0.20$ & 1.72$\pm$0.10\\
    J2031 & 5218$\pm67$ & 2.66$\pm0.08$ & $-2.28\pm0.13$ & 1.65$\pm$0.06\\
    J2140 & 4855$\pm64$ & 1.44$\pm0.12$ & $-3.05\pm0.14$ & 2.02$\pm$0.06\\
    \hline
        \end{tabular}%
    }
\end{table}

Following the parameter determination, the elemental abundances were derived via EW analysis and spectral synthesis. We used $\alpha$-enhanced ($\mathrm{[\alpha/Fe] = +0.4}$) \code{ATLAS9} model atmospheres \citep{castelli2003} and Solar abundances were taken from \citet{asplund2009}. Line lists used for the analysis were generated from \code{linemake}\footnote{\href{https://github.com/vmplacco/linemake}{https://github.com/vmplacco /linemake}} \citep{placco2021}, and include isotopic and hyperfine structure broadening, where applicable, employing the $r$-process isotope ratios from \citet{sneden2008}. Atomic data, EWs, and derived abundances for individual lines used are listed in Table \ref{tab:lines}. 
Final abundances were determined as weighted averages of individual line abundances following \cite{2020AJ....160..181J}. We also follow the procedure outlined in \citet{2020AJ....160..181J} to determine the abundance uncertainties by propagating through the stellar parameter uncertainties (see Table \ref{table:uncertainties}).

\section{Results}\label{results}
Abundances of 30 elements, including 10 neutron-capture elements, were determined for the three stars. Final abundances and associated uncertainties are listed in Table \ref{table:abundances}. In Table \ref{table:r-lim}, we list the abundance ratios for the three stars associated with the $r_{lim}$ abundance criteria. Figure \ref{fig:all_elements} compares the derived abundances for selected elements to those of normal Milky Way (MW) halo stars (black circles) from \citet{Roederer2014} and $r_{lim}$ stars (red stars) from the literature. The sample of literature $r_{lim}$ stars was compiled from the SAGA Database \citep{2008PASJ...60.1159S,2011MNRAS.412..843S,2013MNRAS.436.1362Y,2017PASJ...69...76S}, selected so that they fulfil the criteria of $r_{lim}$ stars (see Table \ref{table:r-lim}). We only included stars that had measured abundances for all three -- namely Sr, Ba and Eu -- elements, and excluded those where only upper limits were available. Abundances from the following studies are included in Figure \ref{fig:all_elements}: \citet{2005A&A...439..129B,2006AJ....132...85P,2007A&A...476..935F,2008ApJ...681.1524L,2013ApJ...778...56C,2013ApJ...771...67I,hansen2018,2018ApJ...868..110S,2020ApJ...898..150E,holmbeck2020}.

\subsection{Light elements Li to Zn}
\begin{figure*}
    \centering
\includegraphics[width=\textwidth]{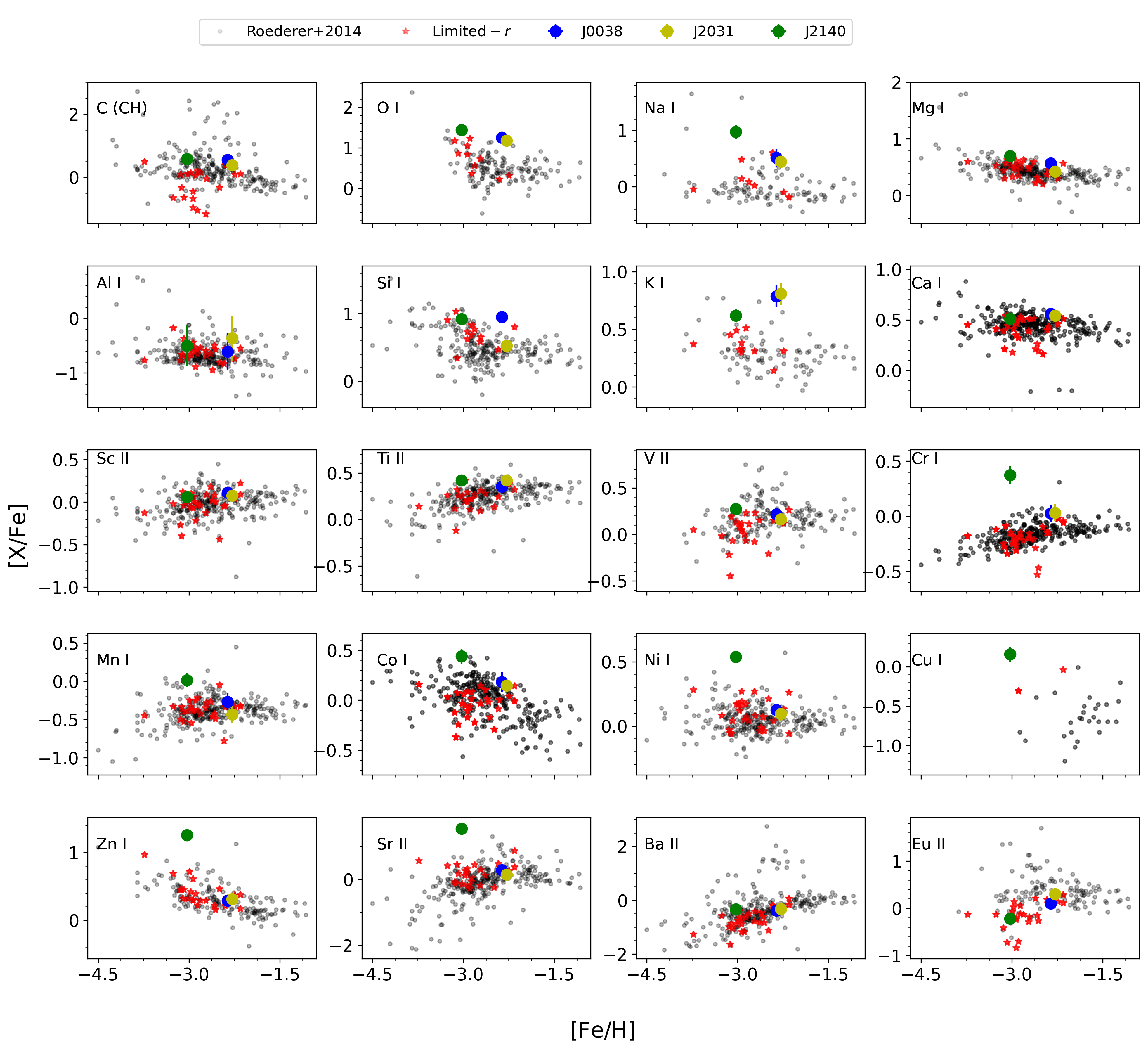}
    \caption{Derived abundances for the three sample stars (blue, yellow, and green dots) compared to abundances of normal MW halo stars (black dots) from \citet{Roederer2014} and literature $r_{lim}$ stars (red stars). The error bars of our three stars, when not visible, are the size of the dots.
    \label{fig:all_elements}}
\end{figure*}

We derived the abundances of elements from Li to Zn using a combination of EW and spectral synthesis analysis. See Table \ref{tab:lines} for details on the individual lines used. From inspection of Figure \ref{fig:all_elements}, the abundances derived for J0038 and J2031 generally follow the trends seen for other MW halo and $r_{lim}$ stars for the elements displayed, with the exception of O and K. On the other hand, J2140 generally exhibits higher abundances of the iron-peak elements Cr, Mn, Co, Ni, Cu, and Zn, as well as for O and K, similar to J0038 and J2031. This star is also enhanced in N and Na, suggesting it has experienced a different chemical-enrichment history compared to the typical MW halo star and the two other stars in our sample. It is evident in Figure \ref{fig:all_elements} that J2140 stands out from all other stars when looking at the iron-peak and Na abundances. This is also very interesting when taking into account the fact that the $\alpha$-element abundances of J2140 follow the trend of the typical metal-poor MW halo stars, with the exception of O, which is somewhat higher. 

\begin{figure*}
    \centering
\includegraphics[scale=0.4]{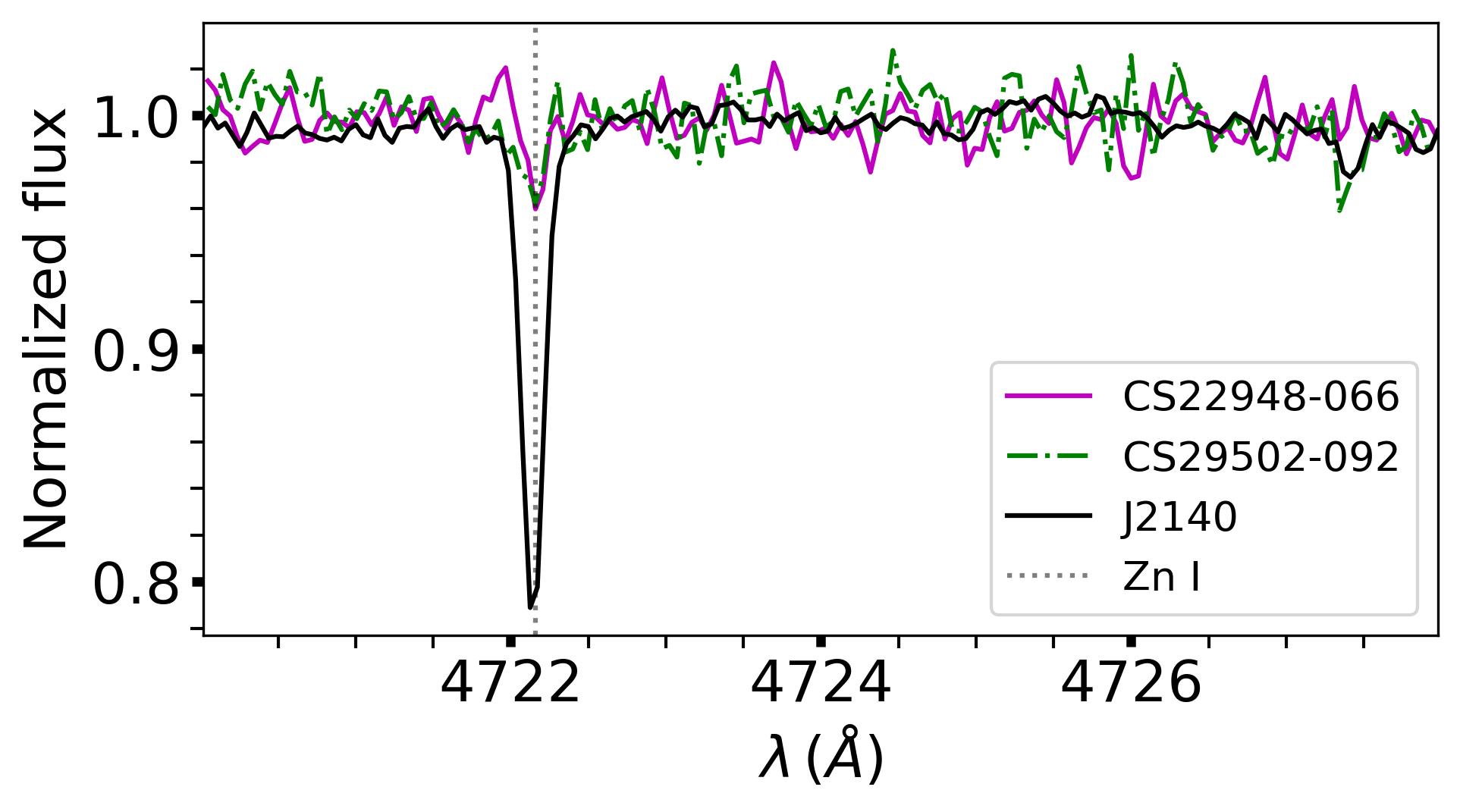}
\includegraphics[scale=0.4]{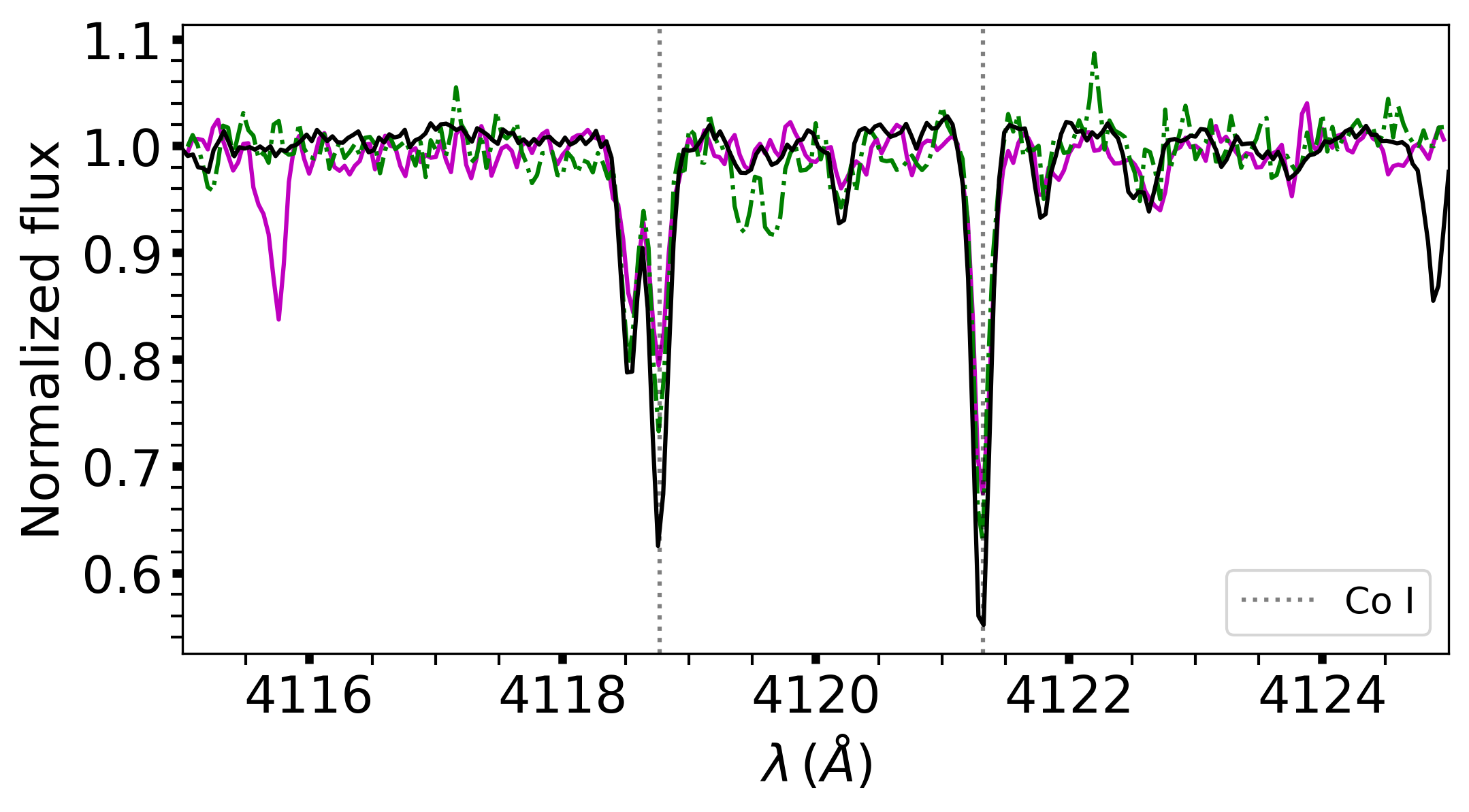}
\includegraphics[scale=0.4]{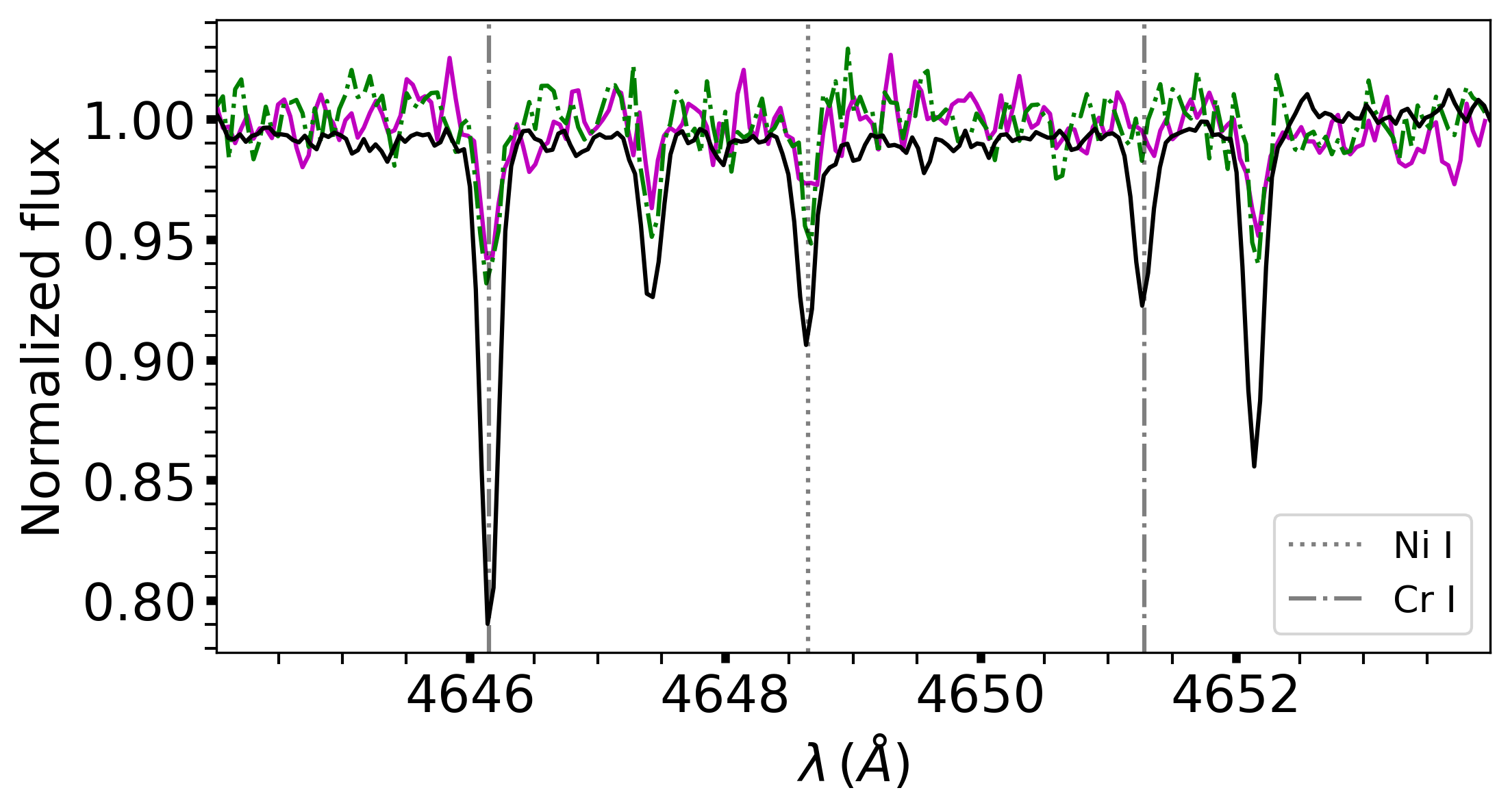}
\includegraphics[scale=0.4]{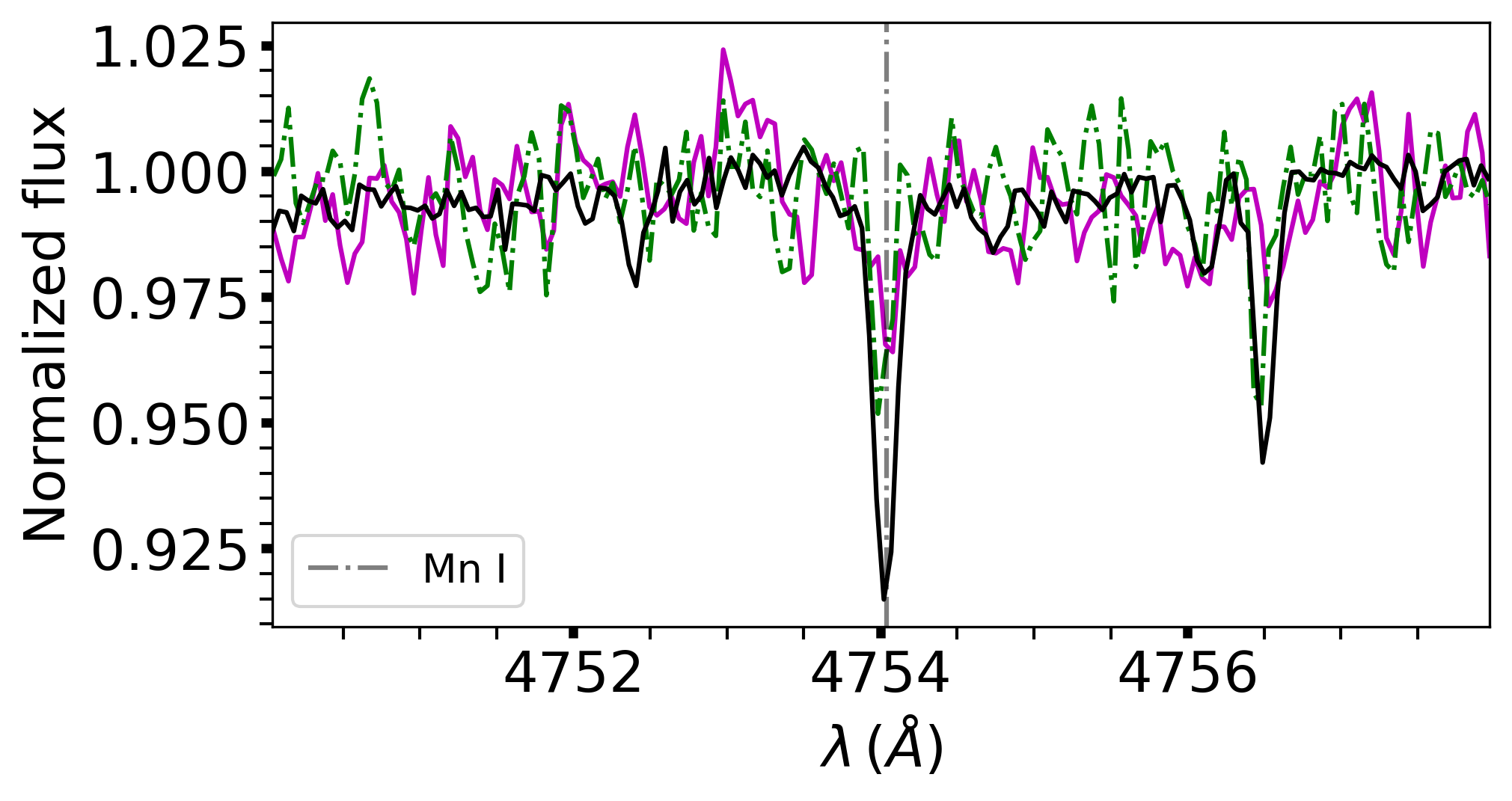}
    \caption{Comparison of spectral lines between J2140 and stars CS~29502$-$092 and CS~22948$-$066. The spectral lines of \ion{Zn}{i} at 4722.16 \AA\, (top left panel), of \ion{Co}{i} at 4118.77 \AA\, and 4121.32 \AA\, (top right panel), of \ion{Ni}{i} and \ion{Cr}{i} (bottom left panel) at 4604.99 \AA\, and 4648.65 \AA, and 4646.15 \AA\, and 4651.28 \AA, respectively, and of \ion{Mn}{i} at 4754.04 \AA\, (bottom right panel) are shown.}
    \label{fig:comparison} 
\end{figure*}

In Figure \ref{fig:comparison}, we compare some spectral lines of Cr, Mn, Co, Ni, and Zn from J2140 to those of two other stars with similar stellar parameters, in order to demonstrate the enhancement this star exhibits in those elements. The comparison stars are CS~29502$-$092 and CS~22948$-$066 with stellar parameters $\mathrm{T_{eff}}$=4820$\pm$34 K, $\mathrm{\log g}$=1.5$\pm$0.14, and $\mathrm{[Fe/H]}=-3.2\pm0.15$, and $\mathrm{T_{eff}}$=4830$\pm$34 K, $\mathrm{\log g}$=1.55$\pm$0.15, and $\mathrm{[Fe/H]}=-3.18\pm0.16$, respectively \citep{Roederer2014}. The absolute abundances of those stars that were reported by \cite{Roederer2014}, are: $\log \,\epsilon(\mathrm{Cr I})=2.19$, $\log \,\epsilon(\mathrm{Mn I})=2.11$, $\log \,\epsilon(\mathrm{Ni I})=3.21$,	$\log \,\epsilon(\mathrm{Co I})=1.73$, $\log \,\epsilon(\mathrm{Zn I})=1.70$ for CS29502-092, and $\log \,\epsilon(\mathrm{Cr I})=1.82$, $\log \,\epsilon(\mathrm{Mn I})=1.85$, $\log \,\epsilon(\mathrm{Ni I})=2.86$,	$\log \,\epsilon(\mathrm{Co I})=1.67$, $\log \,\epsilon(\mathrm{Zn I})=1.58$ for CS22948-066. 

We used the non-local thermodynamic equilibrium (NLTE) corrections from \cite{Bergemann2021} for \ion{O}{i}, and from \cite{Andrievsky_2010} for \ion{K}{i} in order to assess whether the over-abundances of those elements for all three stars are merely NLTE effects, rather than real enhancements. We did the same for \ion{Cr}{i}, \ion{Mn}{i}, and \ion{Co}{i} only for J2140 using the NLTE corrections from \cite{Bergemann2010b}, \cite{Bergemann2019}, and from \cite{Bergemann2010} with collisional data from \cite{Voronov_2022}, respectively. In the case of \ion{O}{i}, the NLTE corrections for J0038 and J2031 are on the order of $\sim-0.03$ dex, whereas no correction arises for J2140. Regarding \ion{K}{i}, the NLTE corrections are $\sim\,-0.21$ dex for J2140, and $\sim\,-0.27$ dex for J0038 and J2031. Concerning \ion{Cr}{i}, \ion{Mn}{i}, and \ion{Co}{i}, the NLTE corrections for J2140 are $\sim\,+0.55$ dex, $\sim\,+0.4$ dex, and $\sim\,+0.87$ dex, respectively. Finally, after application of the evolutionary correction from \cite{Placco2014}, J2140 has $\mathrm{[C/Fe]}=+1.05$, which would classify it as a carbon-enhanced metal-poor (CEMP) star \citep{beers2005,Aoki2007,Carollo_2012,Norris_2013}.

\subsection{Neutron-capture elements}
\begin{figure*}
    \centering
\includegraphics[scale=0.45]{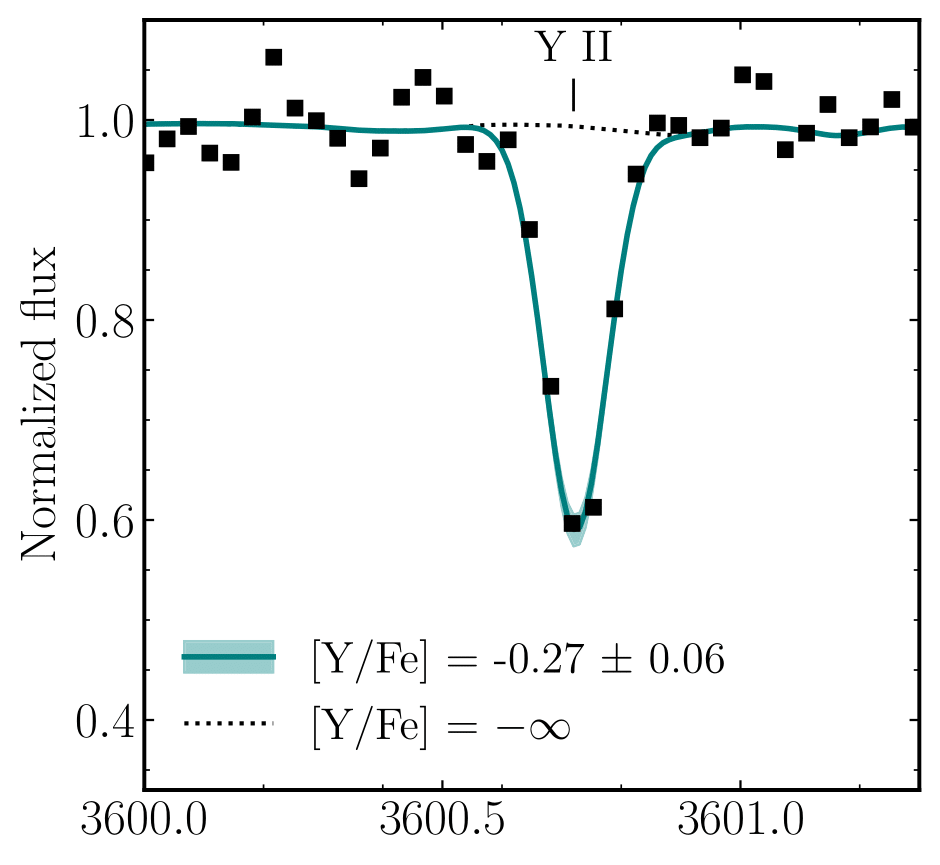}
\includegraphics[scale=0.45]{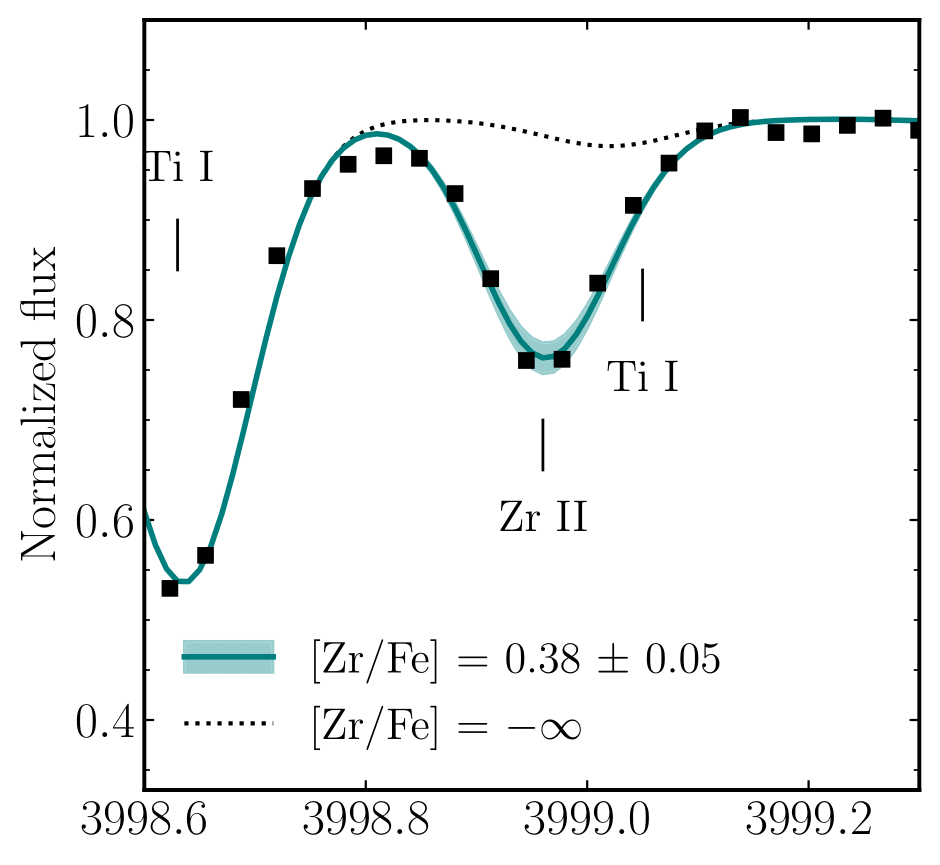}
\includegraphics[scale=0.45]{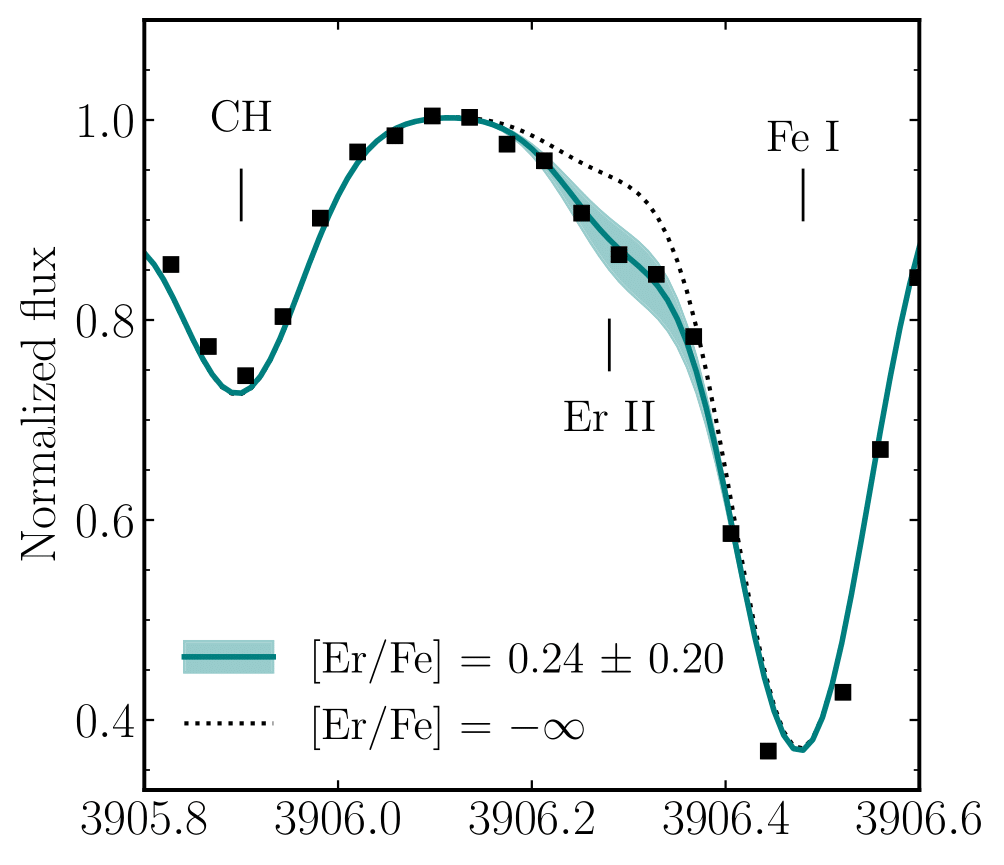}
    \caption{Comparison of synthesis and observed spectra (black dots) for an \ion{Y}{ii} line in J2031 (left panel), a \ion{Zr}{ii} line in J0038 (middle panel), and an \ion{Er}{ii} line in J2140 (right panel). The blue line is the best-fit synthesis, the blue band shows the uncertainty, and the dotted line is a synthesis without the given element. As demonstrated, the spectral synthesis technique we employed can reproduce very well the observed single (\ion{Y}{ii}), blended (\ion{Zr}{ii}), and weak (\ion{Er}{ii}) lines.}
    \label{fig:synthesis} 
\end{figure*}

We derived abundances of ten neutron-capture elements, specifically Sr, Y, Zr, Ba, La, Nd, Eu, Dy, Er, Yb, via spectral synthesis. Synthesis of neutron-capture element absorption features present in the spectra of the three stars are shown in Figure \ref{fig:synthesis}. In Table \ref{table:r-lim}, we list the $\mathrm{[Eu/Fe]}$, $\mathrm{[Sr/Ba]}$, and $\mathrm{[Sr/Eu]}$ ratios for the stars, along with the limits required for a $r_{lim}$ classification according to \citet{frebel2018}. Two of our stars, J0038 and J2140, fulfil the criteria of $r_{lim}$ stars; $\mathrm{[Eu/Fe]}<+0.3$, $\mathrm{[Sr/Ba]}>+0.5$ and $\mathrm{[Sr/Eu]}>0.0$, while J2031 has a too-high  $\mathrm{[Eu/Fe]}$ ratio and a too-low $\mathrm{[Sr/Eu]}$ ratio, and can be classified as an $r$-I star. A somewhat cooler spectroscopic $T_{\rm eff}$ of 4894~K and lower gravity of $\log g = 1.39$ was derived in \cite{hansen2018}, likely resulting in the lower $\mathrm{[Eu/Fe]}$ abundances derived, and subsequent $r_{lim}$ classification of this star. Its $\mathrm{[Eu/Fe]} =+0.3$ abundance also barely makes it qualify for the $r$-I class, and thus, it may be useful for exploring the transition between the $r_{lim}$ and $r$-I regime.

\begin{table}
\caption{Sample of line information.\label{tab:lines}}
\centering
\resizebox{\columnwidth}{!}{%
\begin{tabular}{llrrrrrrr}
\hline \hline
$\mathrm{Stellar ID}$ &  $\mathrm{Species}$ &  $\mathrm{\lambda}$ &  $\mathrm{\chi}$ &  $\mathrm{\log\,}gf$ &  $\mathrm{EW}$& $\sigma_{\mathrm{EW}}$ &$\mathrm{\log \epsilon}$ & $\mathrm{ref}$\\
&  & (\AA) &($\mathrm{eV}$) & & ($\mathrm{m}$\AA)& ($\mathrm{m}$\AA)&&\\\hline
J003859 &     O I &     7771.94 &   9.15 &   0.37 &   20.48 &   2.62 &   7.65 &   1 \\
   J003859 &     O I &     7774.17 &   9.15 &   0.22 &    7.73 &   1.62 &   7.24 &   1 \\
   J003859 &     O I &     7775.39 &   9.15 &   0.00 &    7.87 &    1.53 &   7.47 &   1 \\
   J003859 &    Na I &     5889.95 &   0.00 &   0.11 &  170.38 &   2.93 &   4.21 &   1 \\
   J003859 &    Na I &     5895.92 &   0.00 &  $-$0.19 &  158.03 &   0.78 &   4.36 &   1 \\
   J003859 &    Mg I &     4167.27 &   4.35 &  $-$0.74 &   61.95 &   1.21 &   5.66 &   1 \\
   J003859 &    Mg I &     4702.99 &   4.33 &  $-$0.44 &   80.18 &   0.71 &   5.54 &   1 \\
   J003859 &    Mg I &     5528.40 &   4.35 &  $-$0.55 &   79.94 &   0.72 &   5.68 &   2 \\
   J003859 &    Mg I &     5711.09 &   4.35 &  $-$1.84 &   14.72 &   0.80 &   5.79 &   2 \\
   J003859 &    Al I &     3961.52 &   0.01 &  $-$0.33 &  121.84 &   1.40 &   3.35 &   1 \\
   J003859 &    Si I &     5772.15 &   5.08 &  $-$1.75 &    5.58 &   1.21 &   5.98 &   1 \\
   J003859 &     K I &     7664.90 &   0.00 &   0.12 &   65.09 &   0.89 &   3.37 &   1 \\
   J003859 &     K I &     7698.96 &   0.00 &  $-$0.18 &   45.82 &   0.88 &   3.33 &   1 \\
   \hline
\end{tabular}%
}
\tablebib{(1) \cite{kramida2018aps}, (2) \cite{2017arXiv170806692P}, (3) \cite{2018ADNDT.119..263Y}, (4) \cite{2013ApJS..205...11L}, (5) \cite{2013ApJS..208...27W}, (6) \cite{2001ApJS..132..403P,2002ApJS..138..247P}, (7) \cite{2007ApJ...667.1267S}, (8) \cite{2017ApJS..228...10L}, (9) \cite{2014ApJS..215...23D} , (10) \cite{1991JOSAB...8.1185O}, (11) \cite{2017ApJ...848..125B},  (12) \cite{2014MNRAS.441.3127R}, (13) \cite{2009A&A...497..611M}, (14) \cite{2019ApJS..243...33D}, (15) \cite{2014ApJS..211...20W}, (16) \cite{2012ApJ...750...76R}, (17) \cite{1998ApJ...506..405S}, (18) \cite{1989JOSAB...6.1457L} using hfs from \cite{kurucz1995kurucz}, (19) \cite{Lawler_2014} (20) \cite{2014ApJS..211...20W}, (21) \cite{2011ApJS..194...35D}, (22) \cite{2015ApJS..220...13L}, (23) \cite{Kramida2018b}, (24) \cite{2011MNRAS.414.3350B}, (25) \cite{2006A&A...456.1181L}, (26) \cite{Kramida2018b}  using HFS/IS from \cite{1998AJ....115.1640M}, (27) \cite{2001ApJ...556..452L} using HFS from \cite{2006ApJ...645..613I}, (28) \cite{2003ApJS..148..543D} using HFS/IS from \cite{Roederer_2008}, (29) \cite{Lawler_2001} using HFS/IS from \cite{2006ApJ...645..613I} (30) \cite{2000JQSRT..66..363W}, (31) \cite{2008ApJS..178...71L}, (32) \cite{2009ApJS..182...80S}, (33) \cite{Kramida1999-co} using hfs from \cite{kurucz1995kurucz}. }
\tablefoot{The full table is available online.}
\end{table}

\onecolumn

\begin{table*}
\caption{Abundance table. N denotes the number of absorption lines used for the elemental-abundance determination.}
\label{table:abundances}
\centering
\resizebox{\columnwidth}{!}{%
\begin{tabular}{lrrrrrrrrrrrrrrrrrr}
\multicolumn{19}{l}{\hspace*{2cm} 2MASSJ00385967+2725516 \hspace*{3cm} 2MASSJ20313531-3127319 \hspace*{4cm} 2MASSJ21402305-1227035}\\

\hline \hline
$\mathrm{Element}$ &  $\mathrm{N}$ &  $\log \epsilon (X)$ &  $\mathrm{[X/H]}$ &  $\mathrm{\sigma_{[X/H]}}$ &  $\mathrm{[X/Fe]}$ &  $\mathrm{\sigma_{[X/Fe]}}$ &  $\mathrm{N}$ &  $\log \epsilon (X)$ &  $\mathrm{[X/H]}$ &  $\mathrm{\sigma_{[X/H]}}$ &  $\mathrm{[X/Fe]}$ &  $\mathrm{\sigma_{[X/Fe]}}$&  $\mathrm{N}$ &  $\log \epsilon (X)$ &  $\mathrm{[X/H]}$ &  $\mathrm{\sigma_{[X/H]}}$ &  $\mathrm{[X/Fe]}$ &  $\mathrm{\sigma_{[X/Fe]}}$\\
& &  & &(dex) & &(dex)& &  & &(dex) & &(dex)& &  & &(dex) & &(dex)\\\hline
                  Li I &               1 &             +1.04 &                 $-$0.01 &                           0.10 &                   +2.47 &                            0.10 &               1 &             +1.16 &                  +0.11 &                           0.06 &                   +2.48 &                            0.06 &             - &            - &                 - &                          - &                  - &                           - \\
                  C-H &               1 &             +6.50 &                 $-$1.93 &                           0.12 &                   +0.55 &                            0.11 &               1 &             +6.44 &                 $-$1.99 &                           0.06 &                   +0.38 &                            0.06 &             1 &           +5.88 &               $-$2.55 &                         0.08 &                 +0.58 &                          0.07 \\
                  $\mathrm{C_{cor}}$ &                &              &                  &                            &                   +0.56\text{*} &                             &                &              &                 &                            &                   +0.39\text{*} &                             &             &            &               &                          &                 +1.05\text{*} &                           \\
                  N-H &               - &              - &                   - &                            - &                    - &                             - &               - &              - &                   - &                            - &                    - &                             - &             1 &           +6.04 &               $-$1.79 &                         0.14 &                 +1.33 &                          0.13 \\
                 O I &               3 &             +7.46 &                 $-$1.23 &                           0.10 &                   +1.25 &                            0.12 &               2 &             +7.49 &                 $-$1.20 &                           0.04 &                   +1.18 &                            0.06 &             3 &           +7.00 &               $-$1.69 &                         0.12 &                 +1.44 &                          0.13 \\
                Na I &               2 &             +4.27 &                 $-$1.97 &                           0.17 &                   +0.51 &                            0.16 &               3 &             +4.31 &                 $-$1.93 &                           0.09 &                   +0.44 &                            0.09 &             2 &           +4.08 &               $-$2.15 &                         0.13 &                 +0.97 &                          0.13 \\
                Mg I &               4 &             +5.69 &                 $-$1.91 &                           0.08 &                   +0.57 &                            0.08 &               7 &             +5.65 &                 $-$1.95 &                           0.08 &                   +0.42 &                            0.08 &             8 &           +5.17 &               $-$2.43 &                         0.07 &                +0.70 &                          0.07 \\
                Al I &               2 &             +3.36 &                 $-$3.09 &                           0.34 &                  $-$0.61 &                            0.33 &               3 &             +3.71 &                 $-$2.74 &                           0.42 &                  $-$0.36 &                            0.41 &             1 &           +2.82 &               $-$3.63 &                         0.39 &                $-$0.50 &                          0.38 \\
                Si I &               2 &             +5.98 &                 $-$1.53 &                           0.08 &                   +0.95 &                            0.08 &               4 &             +5.66 &                 $-$1.85 &                           0.04 &                   +0.53 &                            0.05 &             3 &           +5.30 &               $-$2.21 &                         0.06 &                 +0.92 &                          0.06 \\
                 K I &               2 &             +3.34 &                 $-$1.69 &                           0.10 &                   +0.79 &                            0.09 &               2 &             +3.46 &                 $-$1.57 &                           0.10 &                   +0.81 &                            0.10 &             1 &           +2.52 &               $-$2.51 &                         0.05 &                 +0.62 &                          0.05 \\
                Ca I &              24 &             +4.42 &                 $-$1.92 &                           0.06 &                   +0.56 &                            0.06 &              28 &             +4.51 &                 $-$1.83 &                           0.04 &                   +0.54 &                            0.04 &            16 &           +3.72 &               $-$2.62 &                         0.06 &                 +0.51 &                          0.06 \\
               Sc II &               8 &             +0.90 &                 $-$2.25 &                           0.07 &                   +0.11 &                            0.07 &              11 &             +0.94 &                 $-$2.21 &                           0.06 &                   +0.08 &                            0.06 &            12 &           +0.18 &               $-$2.97 &                         0.09 &                 +0.06 &                          0.07 \\
                Ti I &              16 &             +2.88 &                 $-$2.08 &                           0.08 &                   +0.40 &                            0.07 &              18 &             +2.99 &                 $-$1.96 &                           0.06 &                   +0.42 &                            0.06 &            19 &           +2.33 &               $-$2.62 &                         0.07 &                 +0.51 &                          0.07 \\
               Ti II &              25 &             +2.94 &                 $-$2.01 &                           0.05 &                   +0.35 &                            0.05 &              27 &             +3.08 &                 $-$1.87 &                           0.04 &                   +0.42 &                            0.05 &            30 &           +2.34 &               $-$2.61 &                         0.06 &                 +0.42 &                          0.06 \\
                 V I &               2 &             +1.58 &                 $-$2.35 &                           0.05 &                   +0.14 &                            0.06 &               3 &             +1.59 &                 $-$2.34 &                           0.06 &                   +0.03 &                            0.06 &             2 &           +1.01 &               $-$2.92 &                         0.05 &                 +0.20 &                          0.05 \\
                V II &               6 &             +1.79 &                 $-$2.14 &                           0.03 &                   +0.22 &                            0.04 &               8 &             +1.81 &                 $-$2.12 &                           0.03 &                   +0.16 &                            0.05 &            10 &           +1.17 &               $-$2.76 &                         0.06 &                 +0.27 &                          0.06 \\
                Cr I &               8 &             +3.19 &                 $-$2.45 &                           0.08 &                   +0.02 &                            0.08 &               7 &             +3.30 &                 $-$2.34 &                           0.08 &                   +0.03 &                            0.08 &            11 &           +2.89 &               $-$2.75 &                         0.08 &                 +0.37 &                          0.08 \\
               Cr II &               3 &             +3.21 &                 $-$2.43 &                           0.05 &                  $-$0.07 &                            0.06 &               3 &             +3.46 &                 $-$2.18 &                           0.06 &                   +0.10 &                            0.06 &             3 &           +2.96 &               $-$2.68 &                         0.05 &                 +0.35 &                          0.04 \\
                Mn I &               6 &             +2.68 &                 $-$2.75 &                           0.11 &                  $-$0.27 &                            0.11 &               7 &             +2.62 &                 $-$2.81 &                           0.11 &                  $-$0.43 &                            0.11 &             6 &           +2.32 &               $-$3.11 &                         0.08 &                 +0.02 &                          0.09 \\
               Mn II &               - &              - &                   - &                            - &                    - &                             - &               5 &             +2.86 &                 $-$2.57 &                           0.10 &                  $-$0.28 &                            0.10 &             3 &           +2.32 &               $-$3.11 &                         0.06 &                $-$0.08 &                          0.06 \\
                Fe I &             149 &             +5.02 &                 $-$2.48 &                           0.04 &                   +0.00 &                            0.00 &             132 &             +5.12 &                 $-$2.38 &                           0.04 &                   +0.00 &                            0.00 &           120 &           +4.37 &               $-$3.13 &                         0.04 &                 +0.00 &                          0.00 \\
               Fe II &              13 &             +5.14 &                 $-$2.36 &                           0.04 &                  + 0.00 &                            0.00 &              10 &             +5.21 &                 $-$2.29 &                           0.05 &                   +0.00 &                            0.00 &            11 &           +4.47 &               $-$3.03 &                         0.06 &                 +0.00 &                          0.00 \\
                Co I &               6 &             +2.69 &                 $-$2.30 &                           0.10 &                   +0.18 &                            0.10 &              17 &             +2.76 &                 $-$2.23 &                           0.06 &                   +0.15 &                            0.06 &            17 &           +2.30 &               $-$2.69 &                         0.07 &                 +0.44 &                          0.07 \\
                Ni I &              15 &             +3.86 &                 $-$2.36 &                           0.04 &                   +0.12 &                            0.04 &              14 &             +3.94 &                 $-$2.28 &                           0.05 &                   +0.09 &                            0.06 &            19 &           +3.63 &               $-$2.59 &                         0.04 &                 +0.54 &                          0.04 \\
                Cu I &               - &              - &                   - &                            - &                    - &                             - &               - &              - &                   - &                            - &                    - &                             - &             1 &           +1.22 &               $-$2.97 &                         0.09 &                 +0.16 &                          0.09 \\
                Zn I &               2 &             +2.37 &                 $-$2.19 &                           0.07 &                   +0.29 &                            0.07 &               2 &             +2.50 &                 $-$2.06 &                           0.05 &                   +0.31 &                            0.05 &             3 &           +2.69 &               $-$1.87 &                         0.08 &                 +1.26 &                          0.08 \\
                Sr I &               - &              - &                   - &                            - &                    - &                             - &               - &              - &                   - &                            - &                    - &                             - &             1 &           +1.01 &               $-$1.86 &                         0.05 &                 +1.27 &                          0.05 \\
               Sr II &               2 &             +0.78 &                 $-$2.09 &                           0.11 &                   +0.28 &                            0.12 &               2 &             +0.73 &                 $-$2.14 &                           0.17 &                   +0.15 &                            0.15 &             3 &           +1.38 &               $-$1.49 &                         0.06 &                 +1.54 &                          0.06 \\
                Y II &              11 &            $-$0.23 &                 $-$2.44 &                           0.06 &                  $-$0.08 &                            0.06 &              12 &            $-$0.35 &                 $-$2.56 &                           0.06 &                  $-$0.27 &                            0.06 &            17 &          $-$0.13 &               $-$2.34 &                         0.04 &                 +0.69 &                          0.07 \\
               Zr II &               8 &             +0.60 &                 $-$1.98 &                           0.04 &                   +0.38 &                            0.05 &               9 &             +0.55 &                 $-$2.04 &                           0.04 &                   +0.25 &                            0.05 &            20 &           +0.54 &               $-$2.04 &                         0.06 &                 +0.99 &                          0.06 \\
               Ba II &               5 &            $-$0.56 &                 $-$2.74 &                           0.10 &                  $-$0.38 &                            0.09 &               5 &            $-$0.43 &                 $-$2.61 &                           0.10 &                  $-$0.32 &                            0.08 &             5 &          $-$1.18 &               $-$3.36 &                         0.13 &                $-$0.33 &                          0.10 \\
               La II &               3 &            $-$1.28 &                 $-$2.38 &                           0.10 &                  $-$0.02 &                            0.10 &               2 &            $-$1.14 &                 $-$2.24 &                           0.10 &                   +0.04 &                            0.12 &             1 &          $-$1.77 &               $-$2.87 &                         0.29 &                 +0.16 &                          0.25 \\
               Nd II &               2 &            $-$0.84 &                 $-$2.26 &                           0.06 &                   +0.10 &                            0.06 &               2 &            $-$0.56 &                 $-$1.98 &                           0.09 &                   +0.31 &                            0.09 &             3 &          $-$1.55 &               $-$2.97 &                         0.08 &                 +0.06 &                          0.06 \\
               Eu II &               3 &            $-$1.74 &                 $-$2.26 &                           0.06 &                   +0.10 &                            0.06 &               3 &            $-$1.47 &                 $-$1.99 &                           0.05 &                   +0.30 &                            0.05 &             2 &          $-$2.73 &               $-$3.25 &                         0.07 &                $-$0.22 &                          0.08 \\
               Dy II &               2 &            $-$1.32 &                 $-$2.42 &                           0.13 &                  $-$0.06 &                            0.13 &               2 &            $-$0.94 &                 $-$2.04 &                           0.10 &                   +0.25 &                            0.10 &             - &            - &                 - &                          - &                  - &                           - \\
               Er II &               1 &            $-$1.18 &                 $-$2.10 &                           0.14 &                   +0.26 &                            0.14 &               2 &            $-$1.01 &                 $-$1.93 &                           0.11 &                   +0.36 &                            0.10 &             1 &          $-$1.88 &               $-$2.80 &                         0.24 &                 +0.24 &                          0.20 \\
               Yb II &               1 &            $-$1.63 &                 $-$2.47 &                           0.23 &                  $-$0.11 &                            0.22 &               1 &            $-$1.20 &                 $-$2.04 &                           0.12 &                   +0.24 &                            0.10 &             1 &          $-$2.45 &               $-$3.29 &                         0.18 &                $-$0.26 &                          0.14 \\

               \hline
\end{tabular}%
}
\tablefoot{\text{*} C abundance after the evolutionary correction from \cite{Placco2014}. }
\end{table*}
\twocolumn

\begin{table}
        \caption{Limited-$r$ classification criteria.}
        \label{table:r-lim}
    \centering
    \begin{tabular} {c c c c c}
    \hline\hline
    Object & $\mathrm{[Eu/Fe]}$ & $\mathrm{[Sr/Ba]}$ & $\mathrm{[Sr/Eu]}$ & $\mathrm{[Ba/Eu]}$\\
    \hline\\
    $r_{lim}$ & $<+0.3$ & $>+0.5$ & $>0.0$ &$\cdots$ \\
    \hline
    J0038 & $+0.10$ & $+0.66$ & $+0.18$ & $-0.48$ \\
    J2031 & $+0.30$ & $+0.47$ & $-0.15$ & $-0.62$ \\
    J2140 & $-0.22$ & $+1.87$ & $+1.76$ &$-0.11$ \\
    
    \hline
    
    \end{tabular}%
\end{table}
\section{Discussion}\label{discussion}
\subsection{$R$-process patterns for limited-$r$ stars}
The classical way to analyse the abundance patterns of RPE stars is to compare them to the scaled Solar System $r$-process abundance pattern, as the pattern of heavy $r$-process elements (Ba to Hf) has been observed time and time again to exhibit a universality consistent with the scaled residual $r$-process Solar pattern \citep{sneden2008,Cowan_2021}. However, as described in Section \ref{sec:intro}, when scaling to Eu, this universality does not extend to the light elements (32 < Z  < 56), and also does not seem to apply to $r_{lim}$ stars like HD~122563 \citep{2006ApJ...643.1180H,2007ApJ...666.1189H}, suggesting that a limited $r$-process or neutron-poor $r$-process could be in operation.

Recently, this picture has been challenged by the RPA in the paper by \citet{Roederer_2022}, who investigated the spread in the abundances of eight stars from the literature with varying $r$-process enrichment ($-0.22\leq\mathrm{[Eu/Fe]}\leq+1.32$). However, instead of scaling the full pattern to Eu as is usually done, \citet{Roederer_2022} scaled the light $r$-process elements (Se to Te) to Zr, and only the elements from Ba and up, to Eu. They found that, even though the light $r$-process elements exhibit variations compared to the heavy ones, they are not entirely decoupled. Further, by scaling to Zr, a universal pattern amongst the light $r$-process elements Se, Sr, Y, Zr, Nb, Mo, and Te appeared. However, for some elements, Ru, Rh, Pd, and Ag, the star-to-star scatter persisted. \cite{roederer2023} investigated this scatter further, and found that in RPE stars the abundances of Ru, Rh, Pd, and Ag are correlated to those of heavy $r$-process elements with $63\leq Z \leq 78$, something that is not observed for the neighbouring elements with $34\leq Z \leq 42$ and $48\leq Z \leq 68$. In order to explain this finding, \cite{roederer2023} proposed that these correlations appear due to fission-fragment depositions. Specifically, they assembled metal-poor stars from the literature with $\mathrm{[Ba/Eu]}<-0.3$, to ensure that the $r$-process is the main channel of heavy-element production, and then constructed a pattern of the mean neutron-capture element abundances of stars with $\mathrm{[Eu/Fe]}\leq+0.3$, including the $r_{lim}$ star HD~122563. This so-called baseline pattern is assumed to represent an $r$-process where no fission has occurred. \cite{roederer2023} found that the $r$-process abundance variations of the other stars in the sample that have $\mathrm{[Eu/Fe]}>+0.3$ can be explained by co-production of the $r$-process and fission-fragment depositions of transuranic nuclei, and that this mechanism not only alters the pattern around Ru - Ag, but also for the heavier elements in the regions $64\leq Z \leq78$. This co-production of certain light and heavy $r$-process elements was previously shown by \cite{2020ApJ...896...28V}. \cite{2020ApJ...896...28V} applied the Finite Range Liquid Drop Model (FRLDM) \citep{PhysRevC.101.054607} fission yields on neutron-rich merger ejecta simulations, and found that the late-time fission fragments are deposited at the region around Ru - Ag,  leading up and into the lanthanides, and that this process is the one that most influences the final abundance distribution in this regions. \cite{PhysRevC.103.025806} also found that neutron-rich ejecta in NSMs produce fission fragments that contribute almost entirely to the final abundances of nuclei with $100\leq A \leq180$ (Ru to -- and including -- the lanthanides). However, when the ejecta are less neutron-rich and weak interactions are taken into account, \cite{PhysRevC.103.025806} found that the fission fragments deposit in the region $ A=140-180$, namely the lanthanides.  Finally, it should be noted that the intermediate neutron-capture process ($i$-process; \citealt{1977ApJ...212..149C}) could also contribute to the abundances of the light neutron-capture elements with $32 \leq Z \leq 55$ \citep{2016ApJ...821...37R}.

\begin{figure}[h]
    \centering
\includegraphics[width=0.4\textwidth]{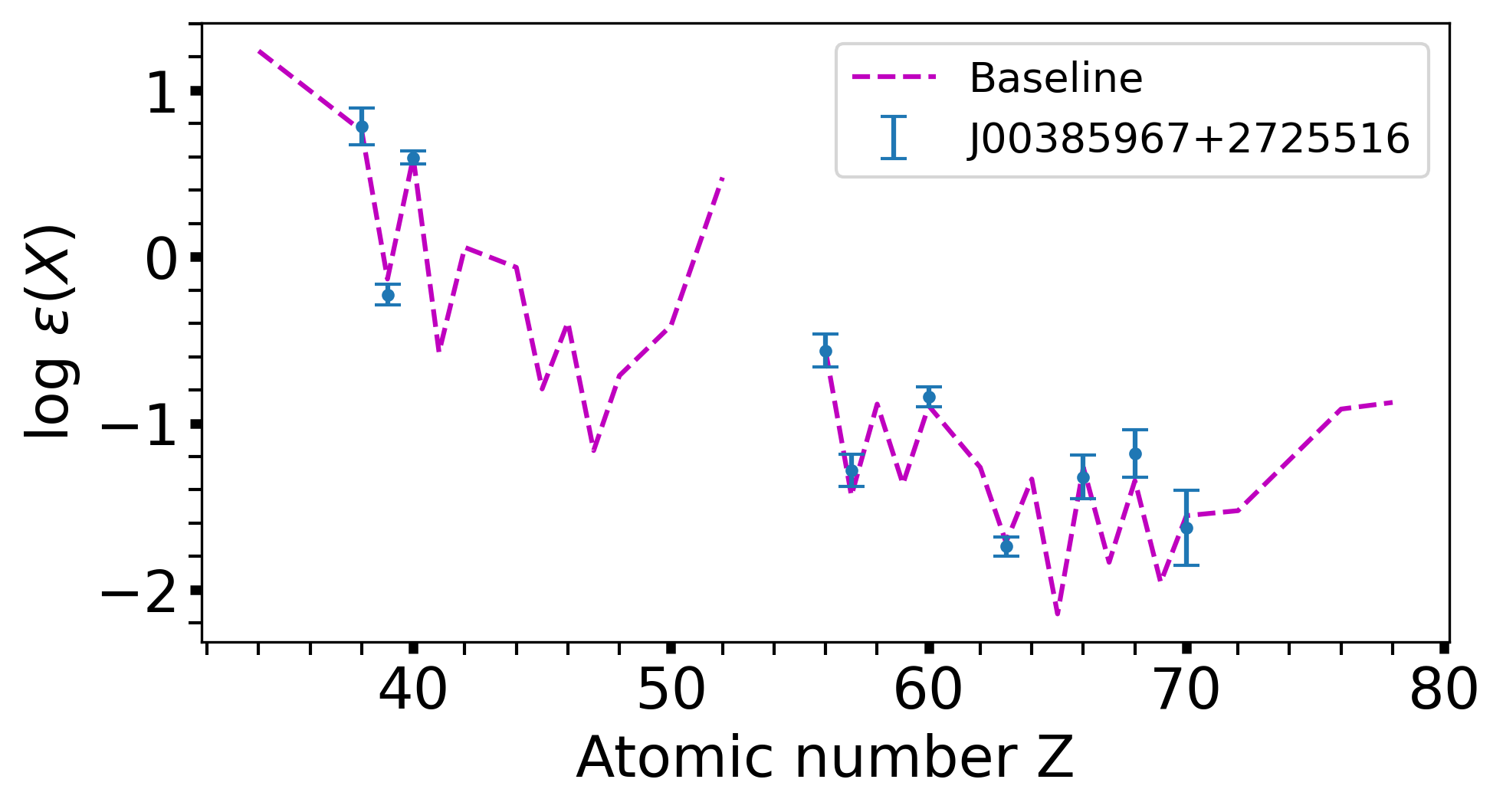}
\includegraphics[width=0.4\textwidth]{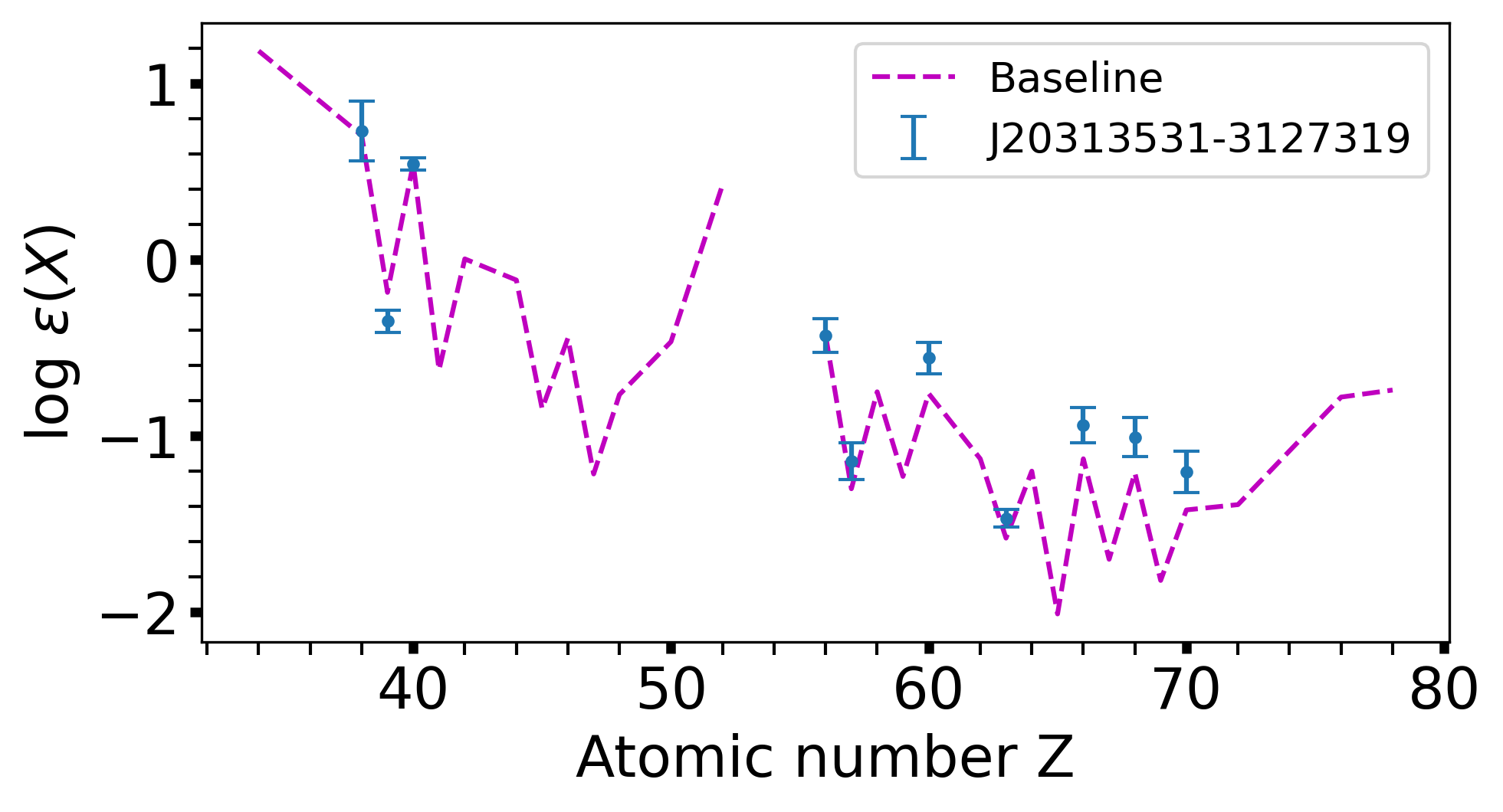}
\includegraphics[width=0.4\textwidth]{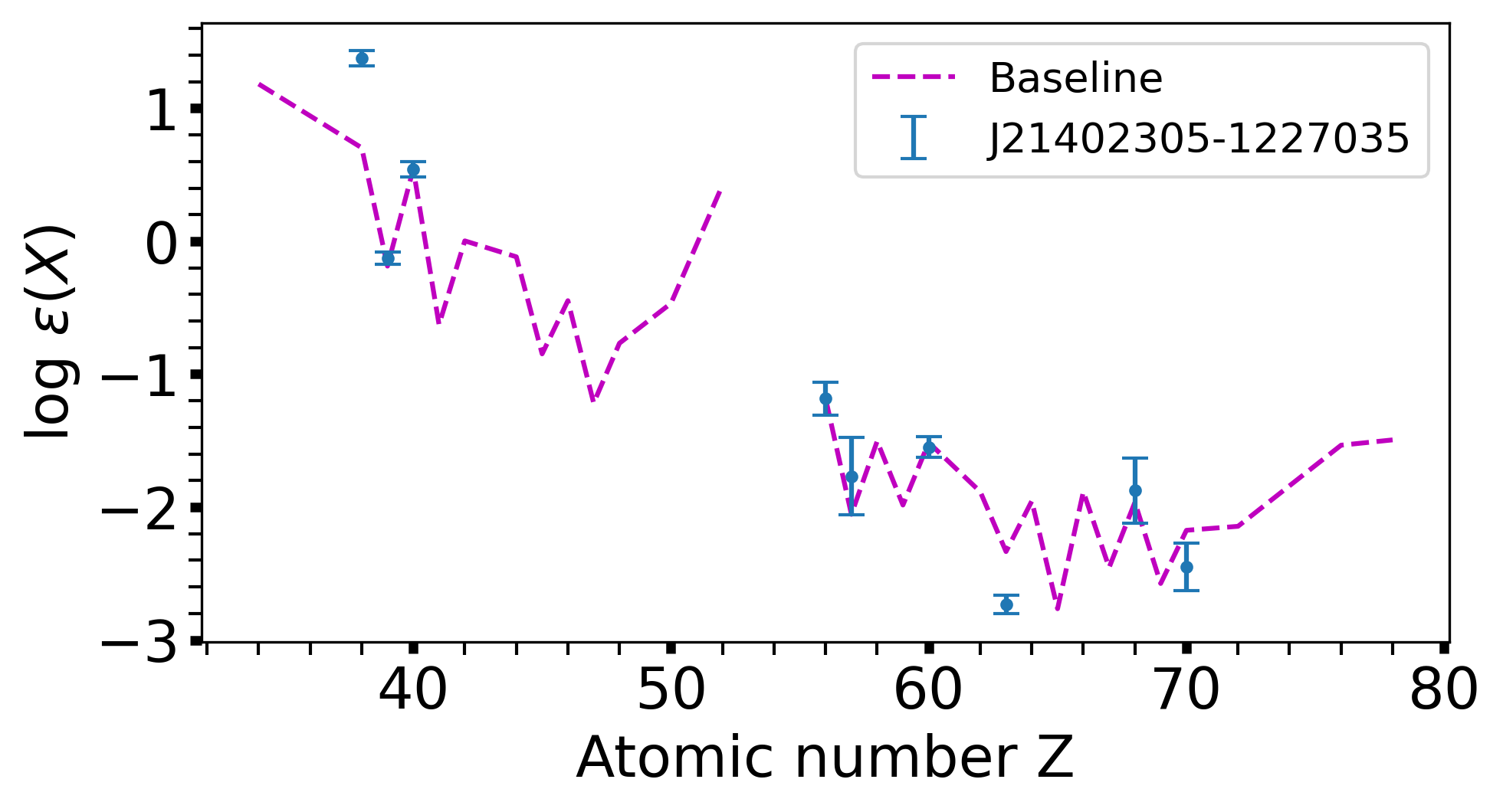}
    \caption{Comparison to the baseline pattern. We plot the abundances of neutron-capture elements for the stars J0038, J2031, and J2140 and over-plot the scaled baseline pattern \citep{roederer2023}. The light elements ($Z<56$) are scaled to Zr, while the heavy ($Z\geq56$) elements are scaled to Ba.}
    \label{fig:baseline_pattern} 
\end{figure}

Figure \ref{fig:baseline_pattern} compares the neutron-capture elemental abundances of our three stars with the baseline pattern from \cite{roederer2023}. In order to do that, we scale the light $r$-process elements to Zr and the heavy ones, that is, $Z\geq56$, to Ba. We find that the $r_{lim}$ star J0038 (top panel) has an abundance pattern that agrees with the baseline pattern very well, suggesting this star could have been enriched by a similar $r$-process as $r$-I and $r$-II stars, but without fission-fragment deposition. The low $\mathrm{[Ba/Eu]}$ ratio for this star of $-$0.48 is also in agreement with that of the stars \cite{roederer2023} used to construct the baseline pattern. J2031 (middle panel), which is an $r$-I star, matches the baseline pattern reasonably well for the light elements (Sr, Y, Zr) but, appears to be more enhanced in some of the heavy elements, suggesting it was enriched by an $r$-process that experienced some fission cycling.However, J2140 (bottom panel) -- which also fulfills the $r_{lim}$ abundance criteria (see Table \ref{table:r-lim}) -- exhibits a somewhat higher Sr abundance and much lower Eu abundance than indicated from the baseline pattern. This could suggest that the heavy elements present in the atmosphere of this star are the products of different or multiple nuclear processes. Since J2140 has $\mathrm{[Ba/Eu]}=-0.11$, some contribution from the $s$-process is likely present, for example from rotating massive stars (spinstars) \citep{2006A&A...447..623M,10.1093/mnras/stv2723,2018ApJS..237...13L}.  In the models of \cite{10.1093/mnras/stv2723}, spinstars can produce elements up to Ba, which are ejected via stellar winds, while the SN models of \cite{2018ApJS..237...13L} including rotation find, that heavier elements up to Pb can be produced.
In principle, spinstars could also contribute to the Sr-Zr abundances we find for J0038 and J2031, however, with their low $\mathrm{[Ba/Eu]}$ values ($-$0.48 and $-$0.62, respectively) and good match to the baseline pattern for Sr-Zr, an $r$-process is more likely. However, the neutron-capture elements in the very old stars are probably formed through the $r$-process, as first suggested by \cite{1981A&A....97..391T}.

\begin{figure}[h]
    \centering
\includegraphics[width=0.4\textwidth]{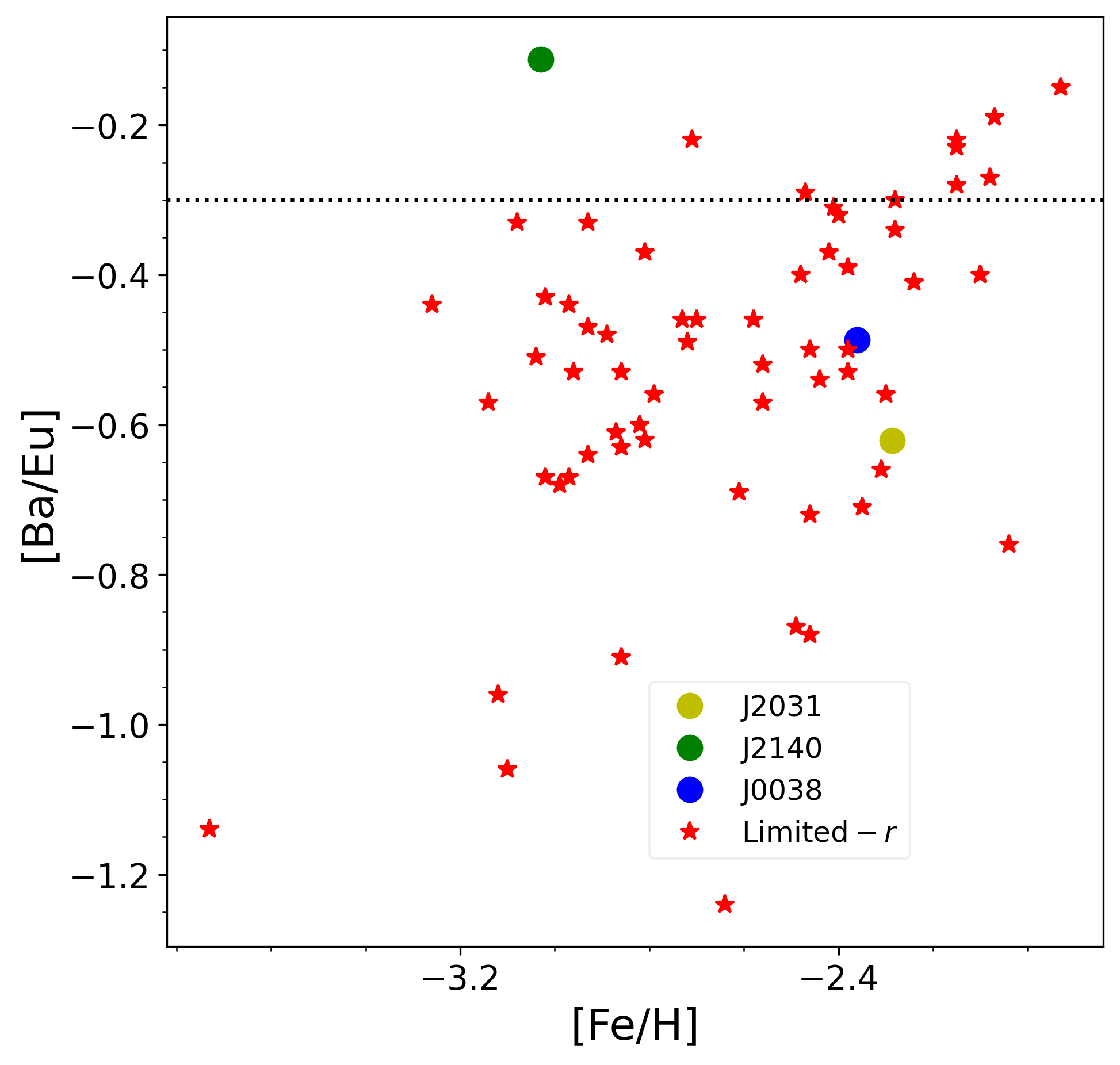}
    \caption{$\mathrm{[Ba/Eu]}$ abundance ratios of our sample of stars and of $r_{lim}$ stars in the literature. Markers are as in Figure \ref{fig:all_elements}. The black-dotted line indicates $\mathrm{[Ba/Eu]}=-0.3$.}
    \label{fig:Ba_Eu_FeH} 
\end{figure}

Based on this comparison, we suggest that, in order to better study the $r_{lim}$ stars, one would also need to take into account the $\mathrm{[Ba/Eu]}$ to be able to distinguish between stars that follow the baseline pattern and those that do not. In Figure \ref{fig:Ba_Eu_FeH} we show the $\mathrm{[Ba/Eu]}$ ratios as a function of $\mathrm{[Fe/H]}$ for our sample stars and the literature $r_{lim}$ stars. Even though $\mathrm{[Ba/Eu]}$ has not been a selection criterion so far for the categorisation of $r_{lim}$ stars, most of them in the literature have $\mathrm{[Ba/Eu]}<-0.3$. Based on the good match between the abundance pattern of star J0038 and the baseline pattern, and the lack of it for star J2140, we believe that studying $r_{lim}$ stars in two regimes, that is $\mathrm{[Ba/Eu]} < -0.3$ and $\geq -0.3$, would help us further understand the formation of these elements.

\subsection{Lanthanide fractions}
Since the $r_{lim}$ stars are selected to exhibit higher abundances in light $r$-process elements compared to the heavy ones, measuring and comparing the ratio of the bulk of light and heavy elements gives us some type of quantification of this over-abundance, which can be useful for identifying the nucleosynthetic channel responsible for the abundance signature of these stars. Because most of the heavy elements in RPE metal-poor stars that are easy to measure belong to the lanthanides, it is straightforward to use the lanthanide mass fraction of the stars in order to quantify the light-to-heavy-element ratio. The lanthanide fraction ($X_\mathrm{La}$) is the ratio of the mass of the elements belonging to the lanthanides to the mass of all other $r$-process elements.

The multi-messenger observations of the gravitational wave event of the NSM GW170817 and its KN is the only evidence we have so far that $r$-process elements are being synthesised in such an event \citep{Kasen_2017,2017ApJ...850L..37P,2017Sci...358.1570D,2018A&A...615A.132R}. In addition, the lanthanide fraction of a KN is a measurable quantity, because it directly affects the duration and shape of the KN light curve as well as the shape of its spectrum \citep{Kasen_2017}. \cite{Ji_2019} computed the $X_\mathrm{La}$ of $r$-process dominated very metal-poor stars ($\mathrm{[Fe/H]}<-2.3$ and $\mathrm{[Ba/Eu]}<-0.4$), and compared them to the $X_\mathrm{La}$ of the KN AT2017gfo. They found that if this KN is a typical representative of a NSM, then such events cannot be the dominant $r$-process site, since most $r$-I and $r$-II stars are more lanthanide rich than this specific KN.

\begin{figure}[h]
    \centering
\includegraphics[width=0.5\textwidth]{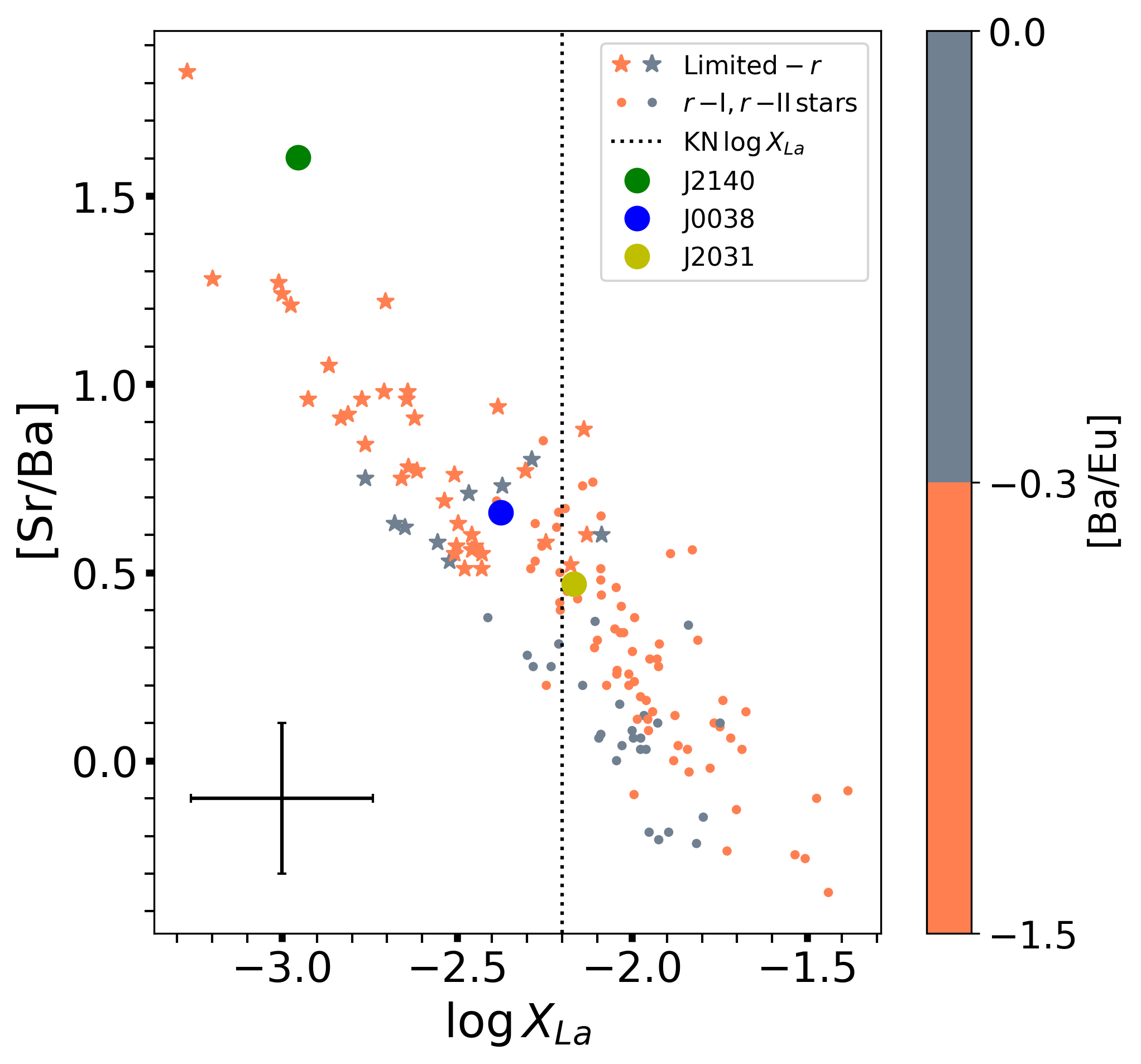}
    \caption{The lanthanide fraction of our sample of stars, and of literature $r_{lim}$, $r$-I, and $r$-II stars. Symbols are as in Figure \ref{fig:all_elements}). Points are coral when $\mathrm{[Ba/Eu]}<-0.3$, and grey when $\mathrm{[Ba/Eu]}\geq-0.3$. The dotted line is the lanthanide fraction of the  KN of the neutron star merger GW170817 \citep{2017Sci...358.1583K,2017ApJ...848L..19C,2017PASJ...69..102T,2017ApJ...848L..27T,2017Natur.551...71T}. The error bar shows the mean uncertainty of the $\mathrm{[Sr/Ba]}$ abundances, and the derived $\log \, X_{La}$'s.}
    \label{fig:lanthanide_fraction}
\end{figure}

We computed the $X_\mathrm{La}$ of our stars, as well as those of the $r_{lim}$ stars in the literature that had abundances measured -- at the very least -- for Sr, Ba, and Eu. To do so, we followed \cite{Ji_2019} and used the Solar residual $r$-process abundances of \cite{sneden2008}. The $X_\mathrm{La}$'s are shown in Figure \ref{fig:lanthanide_fraction}. This result exhibits a clear separation between $r_{lim}$ stars and the $r$-I and $r$-II star around the $X_\mathrm{La}$ value of the KN, which could suggest that the ratio of the light to heavy elements produced in this KN lies in the transition region from $r_{lim}$ to $r$-I, $r$-II stars. This is not surprising since, by design, the selection criteria of $r_{lim}$ stars (Table \ref{table:r-lim}) selects stars with low lanthanide fractions. However, while the $X_\mathrm{La}$ of AT2017gfo might be a good match to $r_{lim}$ stars, the time delay of $r$-process element enrichment by NSMs might cause a problem. As can be seen in Figure \ref{fig:Ba_Eu_FeH}, almost all discovered and analysed $r_{lim}$ stars have $\mathrm{[Fe/H]}<-2.0$, something that has been already discussed by \cite{2019ApJ...875..106C} and \cite{holmbeck2020}. Thus, due to the time delay in the onset of NSMs and the low metallicity of $r_{lim}$ stars, if they indeed bear the imprint of NSMs, they would need to have been born in an environment where star formation is inefficient, which would, in turn, allow the effects of this nucleosynthesis channel to be conspicuous. Recently, however, \cite{2023ApJ...943L..12K} showed that NSMs (including both neutron star (NS)-NS and NS-black hole (BH) mergers) can reproduce the evolutionary relations of $\mathrm{[Eu/Fe]}$-$\mathrm{[Fe/H]}$ and $\mathrm{[Eu/O]}$-$\mathrm{[O/H]}$ in the Solar Neighborhood, when the delay-time distribution (DTD) between onset of star formation and merger is metallicity dependent. An alternative way to eliminate the problem of time delay is to consider MR-SNe or collapsars as a significant source of $r$-process material in the early Universe. \cite{Ji_2019} calculated theoretical $\log \, X_{La}$ values for collapsar models from \cite{Siegel_2019} and MR-SN from \cite{2015ApJ...810..109N}, finding values ranging from $-1.60$ to $-2.81$ and $-0.77$ to $-1.94$ respectively, both overlapping with the value derived for AT2017gfo. In the future, more model calculations and larger stellar samples from the RPA will help determine which sites are dominant. Finally we note, that there might be a bias in the sample of the to date discovered $r_{lim}$ stars, due to the fact that they were discovered in surveys aiming to find stars with $\mathrm{[Fe/H]}<-2.0$.

 \subsection{Kinematics of limited-$r$ stars}
\begin{figure}[h]
    \centering
\includegraphics[width=0.5\textwidth]{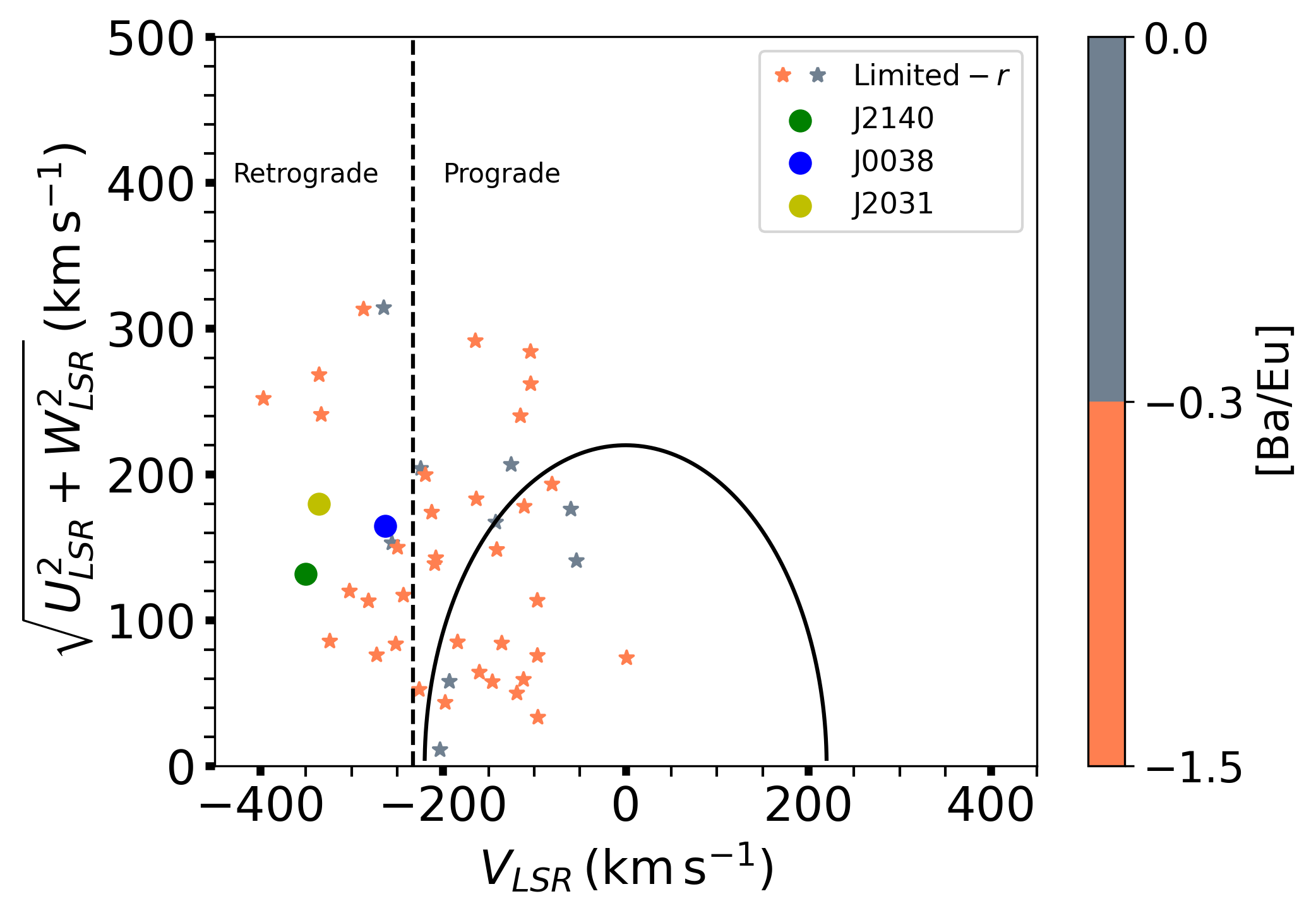}
    \caption{Toomre Diagram for the $r_{lim}$ stars. The velocities depicted were calculated with respect to the LSR. The points are designated as in Figure \ref{fig:lanthanide_fraction}. The dashed black line is $V_{LSR}=-233.1\,\mathrm{km\,s^{-1}}$ \citep{10.1093/mnras/stw2759}, and stars to the left of it are on retrograde orbits. The solid black line designates the area in which all stars have $v_{tot}<220\,\mathrm{km\,s^{-1}}$, were $v_{tot}=\sqrt{U^2+V^2+W^2}$.}
    \label{fig:toomre_lsr}
\end{figure}

 Previous studies have shown that a large fraction of $r$-II stars were likely born in smaller satellite systems and accreted by the MW \citep{roederer2018a,2021ApJ...908...79G,2023ApJ...943...23S}. To investigate if this is also the case for the $r_{lim}$ stars, we used $Gaia$ DR3 radial velocities and proper motions \citep{2023A&A...674A...1G}, and distances from \cite{2021yCat.1352....0B}, to study the kinematics of the $r_{lim}$ stars in our sample and in the literature. The orbits were then calculated with \code{galpy} \footnote{\href{http://github.com/jobovy/galpy}{http://github.com/jobovy/galpy}} \citep{Bovy_2015}. In order to estimate the uncertainties of the orbital parameters, we calculated 500 orbits for each star while varying the proper motions and radial velocities by sampling them from a Gaussian distribution. The distributions had as mean the actual values of the proper motions and radial velocities, while we used their uncertainties as sigma. Figure \ref{fig:toomre_lsr} shows the Toomre Diagram, where we plot $V_{LSR}$ versus $\sqrt{U_{LSR}^2+W_{LSR}^2}$, which are the velocities -- with respect to the Local Standard of Rest (LSR) -- in the Cartesian Galactic coordinate frame. As shown, all three stars from our study have retrograde orbits, suggesting they could have been accreted to the MW from satellite galaxies. However, $\sim$ 65\% of all the $r_{lim}$ stars are on prograde orbits. Also, 38\% of the $r_{lim}$ stars have $v_{tot}<220$ $\mathrm{km\,s^{-1}}$, suggesting they may be consistent with disk stars. In addition, when looking at the $r_{lim}$ stars from the perspective of their $\mathrm{[Ba/Eu]}$ abundance ratio, there also appears to be a difference between the two groups, with most of the stars having $\mathrm{[Ba/Eu]}\geq-0.3$ being on prograde orbits. These findings differ from the findings in the study of \cite{roederer2018a} on the kinematics of 35 highly $r$-process-enhanced field stars ($r$-II for $\mathrm{[Eu/Fe]}>+0.7$). \cite{roederer2018a} showed that most, if not all, of the $r$-II stars, were probably accreted by the MW from ultra-faint dwarf galaxies or low-luminosity dwarf spheroidal galaxies. The study of \citet{roederer2018a} was extended to significantly larger samples by \cite{2021ApJ...908...79G} (466 $r$-I and $r$-II stars) and \cite{2023ApJ...943...23S} (1720 stars). These studies confirm the accreted nature of $r$-I and $r$-II stars. In particular, \cite{2023ApJ...943...23S} finds that only 17\% of the $r$-I stars and 8\% of the $r$-II stars have disk-like kinematics. With the use of an unsupervised learning algorithm, \cite{2023ApJ...943...23S} identified 36 chemo-dynamically tagged groups (CDTGs), and $\sim1\%$ of the $r$-I and $r$-II stars in their sample were identified as belonging to the metal-weak thick disk (MWTD), while $\sim2.1\%$ were traced as members of the splashed disk (SD). The SD is described as a part of the MW primordial disk, which was kinematically heated by the $Gaia$-Sausage-Enceladus (GSE) merger event \citep{2018MNRAS.478..611B,Helmi_2018,DiMatteo2019,Belokurov2020}. 

\begin{figure}[h]
    \centering
\includegraphics[width=0.4\textwidth]{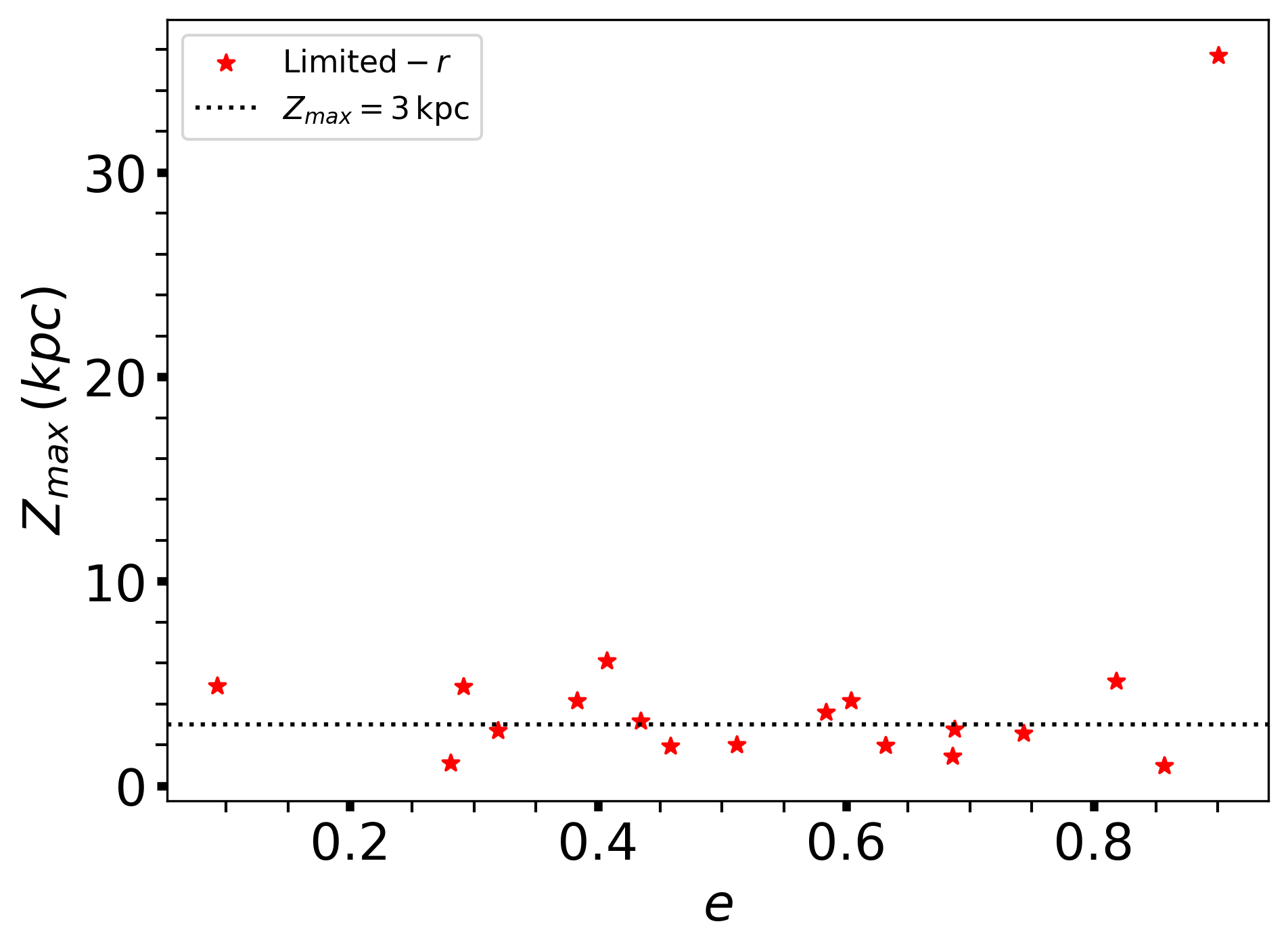}
    \caption{Eccentricity versus $Z_{max}$ of the `disk-like' $r_{lim}$ stars. The black dotted line designates $Z_{max}=3$ kpc. Stars that have $Z_{max}<3$ kpc and $e\leq0.75$ are very likely MWTD stars \citep{Cordoni2020}. The star with the highest eccentricity, $e>0.9$, has also the largest $Z_{max}>30$ kpc.}
    \label{fig:e_zmax}
\end{figure}

\begin{figure}[h]
    \centering
\includegraphics[width=0.4\textwidth]{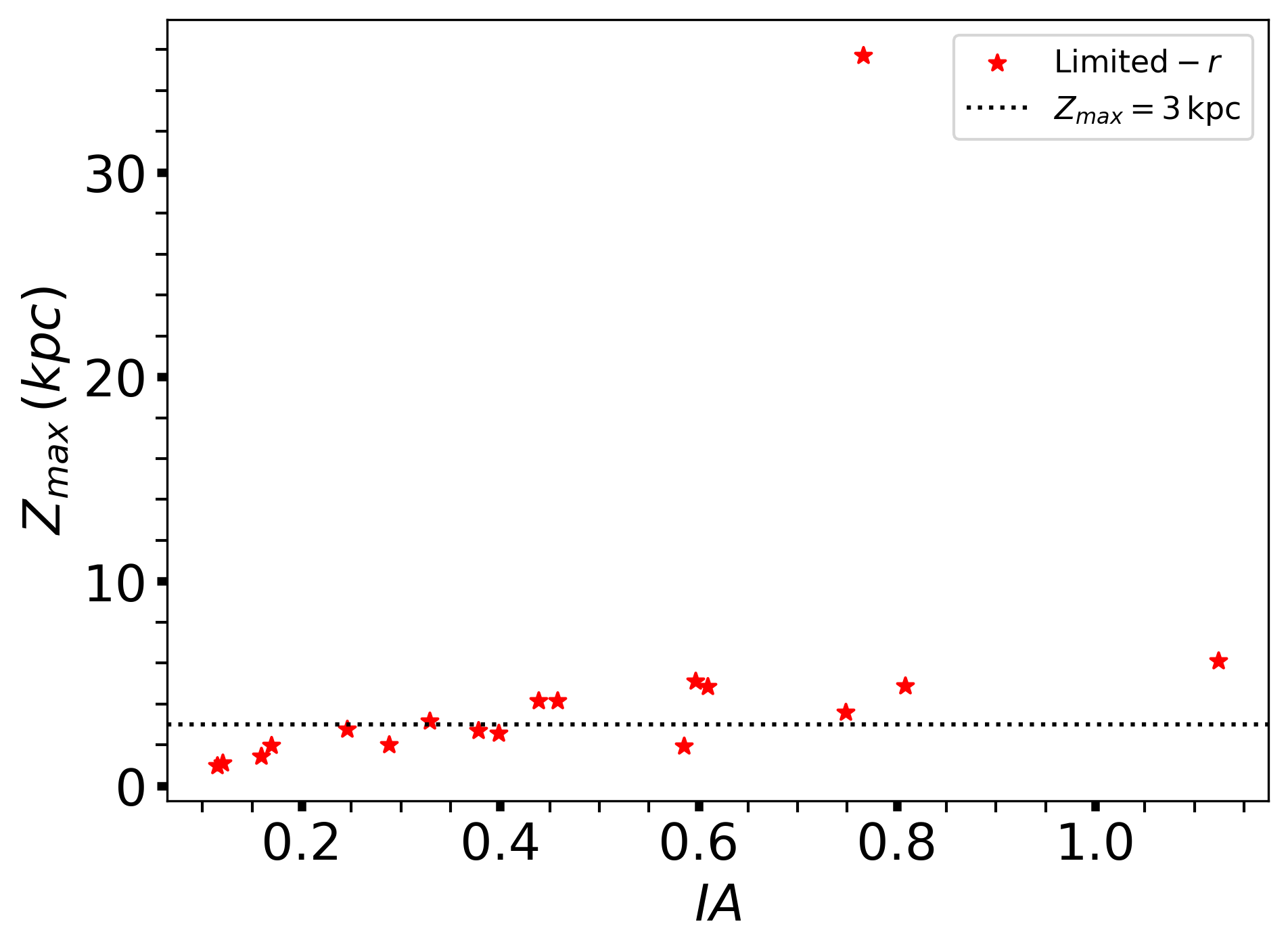}
    \caption{Inclination angle versus $Z_{max}$ of the `disk-like' $r_{lim}$ stars. As in Figure \ref{fig:e_zmax}, the black dotted line designates $Z_{max}=3$ kpc. Stars that have $Z_{max}<3$ kpc or $0.25<IA\leq0.65$ are very likely thick-disk stars, and those with $IA\leq0.25$ are probably thin-disk members \citep{hong2023candidate}. The $IA$ is in radians.}
    \label{fig:IA_zmax}
\end{figure}

 We further examine the $r_{lim}$ stars that appear to be `disk-like', that is, their $v_{tot}=\sqrt{U^2+V^2+W^2}<220\,\mathrm{km\,s^{-1}}$. There have been several studies (for example \citealt{Beers_2014,10.1093/mnras/stz043,Sestito2020,Cordoni2020}) that used the maximum distance of the stars from the Galactic plane, $Z_{max}$, to separate disk from halo stars, often in combination with another orbital parameter. \cite{Cordoni2020} used $Z_{max}$ and the eccentricity of the orbit, $e$, in order to identify disk stars. Specifically, they considered stars on prograde orbits with $\lvert Z_{max} \rvert\leq3$ kpc, and $e<0.75$ to belong to the thick disk. In Figure \ref{fig:e_zmax} we plot $e$ versus $Z_{max}$ of the `disk-like' $r_{lim}$ stars. Based on these criteria, it appears that $\sim39\%$ of the `disk-like' $r_{lim}$ stars belong to the MWTD, which is $\sim15\%$ of all the identified $r_{lim}$ stars (7 stars) to date. Another route to identify disk stars was introduced by \citet{Haywood2018}, who studied stars with high transverse velocities ($v_{t}>200\,\mathrm{km\,s^{-1}}$), and used a $Z_{max}$ - $R_{max}$ plane -- where $R_{max}$ is the apocenter of the orbit projected on the Galactic plane -- and discrete wedges appeared. These wedges were also clearly visible in the distribution of the angles $\arctan(Z_{max}/R_{max})$. Recently, \citet{hong2023candidate} followed \citet{Haywood2018},and assigned ranges to the inclination angle ($IA$) - $IA$=$\arctan(Z_{max}/R_{max})$ - to distinguish thin-, thick-disk, and halo stars. Specifically, they identified stars on prograde orbits as being members of the disk if $\lvert Z_{max}\rvert\leq3$ kpc, or IA$\leq 0.65$. The results following this selection procedure are presented in Figure \ref{fig:IA_zmax}. The use of the $IA$ doubles the percentage of $r_{lim}$ disk stars from $\sim15\%$ to $\sim30\%$ (14 stars). However, 5 out of the 14 stars are --with this criteria-- identified as thin-disk stars ($IA\leq0.25$), while the rest are attributed to the MWTD $0.25<IA\leq0.65$. The significant difference between the $r_{lim}$ and $r$-I, $r$-II stars from the aspect of disk membership still remains, considering that, even though \cite{2023ApJ...943...23S} found 17\% of the $r$-I stars to have disk-like kinematics, only $\sim1\%$ of the $r$-I and $r$-II stars could be chemo-dynamically traced back to the MWTD.

Finally, as our three sample stars all have retrograde orbits, which could point to them being accreted by the MW from a satellite galaxy, we want to investigate their possible association with known structures. The Gaia mission \citep{2016A&A...595A...1G} has to date provided astrometric information for more than a billion stars, enabling astronomers to unravel parts of the hierarchical assembly history of the MW \citep{2020ARA&A..58..205H}. In this context, several accretion events have been identified so far. However, there is not yet a definitive way to select members of an accretion event, in the sense that different kinematic or dynamic selection criteria from different studies can favour different stars as members of the same event, with the existence of a significant overlap. \cite{2021ApJ...908...79G} found that 20\% of the $r$-I and $r$-II stars are connected to the GSE event, while \cite{2023ApJ...943...23S} report 9\% of their sample stars to be associated with this event. We used the dynamic selection criteria from \cite{Myeong_2019} and \cite{Feuillet_2021} for the GSE and Sequoia \citep{Myeong_2019} accretion events. The dynamic criteria of \cite{Myeong_2019} are $-0.07<J_\phi/J_{\mathrm{tot}}<0.07$ and $-1.0<(J_z-J_R)/J_{\mathrm{tot}}<-0.3$ for the GSE event, and $-1.0<J_\phi/J_{\mathrm{tot}}<-0.5$ and $-1.0<(J_z-J_R)/J_{\mathrm{tot}}<0.1$ for the Sequoia event. Those from \cite{Feuillet_2021} are $-500\leq L_z \leq 500$ and $30\leq \sqrt{J_R} \leq 55$, and $-1.0<J_\phi/J_{\mathrm{tot}}<-0.4$ and $-1.0<(J_z-J_R)/J_{\mathrm{tot}}<0.1$, for the GSE and Sequoia events, respectively. According to these criteria, star J2140 was probably accreted during the Sequoia accretion event. In total, two $r_{lim}$ stars seem to be accreted from  $Gaia$ Sequoia, while two to four others were accreted from GSE, depending on the selection criteria. Cumulatively, we find 4-9\% for the $r_{lim}$ stars associated with the GSE, depending on the dynamic criteria employed. The results are shown in Table \ref{table:GSE_Sequoia}.

\section{Summary}\label{summary}
We studied a sample of three $r$-process stars, which were observed by the RPA, and classified as $r_{lim}$ stars. With the updated stellar parameters used for this study, one of the stars, J2031, turned out to be an $r$-I star, while the other two, J0038 and J2140, qualify as $r_{lim}$ stars. The abundances of non-neutron-capture elements for J0038 and J2031 resemble those of normal MW halo stars, while J2140 exhibits higher abundances of the iron-peak elements Cr, Mn, Co, Ni, Cu, and Zn, and is also enhanced in C, N and Na, suggesting that it underwent a different chemical-enrichment history than stars J0038, J2031, and other typical MW halo stars. 

We compared the neutron-capture element abundance patterns of our stars to the baseline pattern of \cite{roederer2023}. For the two $r_{lim}$ stars we find that the pattern of J0038, which has $\mathrm{[Ba/Eu]} < -0.3$, agrees very well with the pattern, while that of J2140, which has $\mathrm{[Ba/Eu]} > -0.3$, does not. This implies that $r_{lim}$ stars with $\mathrm{[Ba/Eu]<-0.3}$ have been enriched by an $r$-process similar to that which enriched $r$-I and $r$-II stars, while another or multiple nuclear processes are responsible for the abundance pattern seen in $r_{lim}$ stars with $\mathrm{[Ba/Eu]} \geq -0.3$. Also, the comparison of the $r$-I star, J2031, with the baseline pattern suggests the abundances of this star have been affected by fission-fragment deposition.

Next, we calculated $X_\mathrm{La}$ of our stars, as well as those of the $r_{lim}$ stars in the literature. We compared it to that of the KN of the NSM GW170817. We find that the $X_\mathrm{La}$ of the KN is in the transition region between $r_{lim}$ stars and $r$-I, $r$-II stars. This could suggest that NSMs like GW170817 could be the $r$-process site responsible for the abundance signatures observed in $r_{lim}$ stars. However, because we do not know the time delay between NSMs and the onset of star formation, it is important to assess whether the $r_{lim}$ stars could have been accreted to the MW from an environment with a low star-formation rate. To investigate that, we studied the kinematics of the $r_{lim}$ stars. We find that, unlike $r$-I and $r$-II stars that were mostly accreted \citep{2021ApJ...908...79G}, 65$\%$ of $r_{lim}$ stars are on prograde orbits suggesting they were probably born in situ. Furthermore, $38\%$ of the $r_{lim}$ stars present disk-like kinematics, which conveys another distinct difference between these and $r$-I, $r$-II stars, as reported by \cite{2023ApJ...943...23S}, who find that 17\% of the $r$-I stars and 8\% of the $r$-II stars have such kinematics. Lastly, we find that 15\% of the $r_{lim}$ stars are simultaneously on prograde orbits, have $Z_{max}\leq3$ kpc, and have $e\leq0.75$, indicating that they belong to the MWTD, unlike the $r$-I, $r$-II stars of which only $\sim1\%$ were chemodynamically attributed to the MWTD \citep{2023ApJ...943...23S}.

The discovery and detailed abundance analysis of more $r_{lim}$ stars is vital for further exploring the kinematic signature of these stars and assessing the difference between those with $\mathrm{[Ba/Eu]}$ above and below $-0.3$. The measurement of additional neutron-capture elements for these stars will either reinforce the fact that the latter seems to have been enriched by an $r$-process similar to that enriching $r$-I and $r$-II stars or provide new insight. Future analysis of snapshot stellar spectra, already obtained by the RPA, is expected to double the number of identified $r_{lim}$ stars.

\begin{acknowledgements}
    We thank the anonymous referee for helpful comments, which helped improve this manuscript. This work was funded by the Deutsche Forschungsgemeinschaft (DFG, German Research Foundation) -- Project-ID 138713538 -- SFB 881 (``The Milky Way System'', subproject A04). T.X. acknowledges support from the Heidelberg Graduate School for Physics (HGSFP). T.T.H. acknowledges support from the Swedish Research Council (VR 2021-05556). T.C.B. acknowledges partial support for this work from grant PHY 14-30152; Physics Frontier Center/JINA Center for the Evolution of the Elements (JINA-CEE), and OISE-1927130: The International Research Network for Nuclear Astrophysics (IReNA), awarded by the US National Science Foundation. R.E. acknowledges support from NSF grant AST-2206263. A.F. acknowledges support from NSF grants AST-1716251 and 2307436. The work of V.M.P. is supported by NOIRLab, which is managed by the Association of Universities for Research in Astronomy (AURA) under a cooperative agreement with the National Science Foundation. I.U.R.\ acknowledges support from the U.S.\ National Science Foundation (NSF), grants PHY~14-30152 (Physics Frontier Center/JINA-CEE), AST~1815403, and AST~2205847, as well as support from the NASA Astrophysics Data Analysis Program, grant 80NSSC21K0627.
    This paper includes data gathered with the 6.5 meter Magellan Telescopes located at Las Campanas Observatory, Chile, and data taken at The McDonald Observatory of The University of Texas at Austin.
    This research made extensive use of the SIMBAD database operated at CDS, Strasbourg, France \citep{wenger2000}, \href{https://arxiv.org/}{arXiv.org}, and NASA's Astrophysics Data System for bibliographic information.
\end{acknowledgements}

\bibliography{references}{}
    \citestyle{egu}
    \bibliographystyle{aa} 

\begin{appendix}
\section{Additional tables.}
We present a table showing the uncertainty arising in the abundance estimation, due to the uncertainty in stellar parameter determination. Further, a table containing information about the likely accreted $r_{lim}$ stars, based on the dynamic selection criteria from \cite{Myeong_2019} and \cite{Feuillet_2021}. Lastly, we include a table that lists the lanthanide fractions and their uncertainties that were computed for the $r_{lim}$ stars.

\begin{table}\label{table:uncertainties}
\caption{Uncertainties in the abundances determination due to the uncertainties in stellar parameters, for the star J0038.}
\centering
\resizebox{\columnwidth}{!}{%
\begin{tabular}{lrrrrr} 
\hline \hline
$\mathrm{Element}$ &  $\Delta T_{\rm eff}$ &  $\mathrm{\Delta \log}g$ &  $\mathrm{\Delta \xi}$ &  $\mathrm{\Delta [M/H]}$ &  $\mathrm{\sigma}_{sys}$ \\
& (K) & (dex) &($\mathrm{km\,s^{-1}}$) &(dex) & (dex)\\\hline
Li I & 0.08 & $-$0.00 & 0.01 & 0.01 & 0.07 \\
C-H  & 0.16 & $-$0.03 & 0.01 & 0.04 & 0.10 \\
O I & $-$0.08 & 0.04 & $-$0.01 & $-$0.00 & 0.04 \\
Na I & 0.10 & $-$0.03 & $-$0.10 & $-$0.00 & 0.16 \\
Mg I & 0.04 & $-$0.01 & $-$0.02 & 0.00 & 0.06 \\
Al I & 0.16 & $-$0.03 & $-$0.11 & 0.01 & 0.23 \\
Si I & 0.04 & 0.01 & $-$0.02 & $-$0.00 & 0.06 \\
K I & 0.06 & $-$0.00 & $-$0.03 & 0.00 & 0.08 \\
Ca I & 0.04 & 0.00 & 0.01 & 0.00 & 0.04 \\
Sc II & 0.02 & 0.03 & $-$0.01 & 0.01 & 0.06 \\
Ti I & 0.08 & $-$0.00 & 0.03 & 0.01 & 0.05 \\
Ti II & 0.03 & 0.03 & 0.04 & 0.02 & 0.03 \\
V I & 0.02 & 0.01 & 0.02 & $-$0.01 & 0.02 \\
V II & $-$0.02 & 0.04 & 0.01 & 0.00 & 0.01 \\
Cr I & 0.08 & 0.00 & 0.03 & 0.01 & 0.05 \\
Cr II & $-$0.00 & 0.03 & $-$0.01 & 0.00 & 0.03 \\
Mn I & 0.06 & $-$0.00 & $-$0.01 & 0.00 & 0.06 \\
Fe I & 0.04 & 0.00 & 0.03 & 0.01 & 0.02 \\
Fe II & 0.01 & 0.03 & 0.02 & 0.01 & 0.02 \\
Co I & 0.05 & 0.01 & $-$0.02 & $-$0.01 & 0.08 \\
Ni I & 0.04 & 0.00 & 0.03 & 0.00 & 0.02 \\
Zn I & 0.04 & 0.01 & $-$0.01 & 0.00 & 0.06 \\
Sr II & $-$0.10 & $-$0.04 & $-$0.19 & $-$0.10 & 0.11 \\
Y II & $-$0.00 & 0.03 & $-$0.00 & 0.01 & 0.04 \\
Zr II & $-$0.02 & 0.04 & 0.01 & 0.00 & 0.02 \\
Ba II & 0.03 & 0.03 & $-$0.02 & 0.01 & 0.08 \\
La II & 0.02 & 0.03 & 0.03 & 0.01 & 0.02 \\
Nd II & 0.02 & 0.02 & 0.01 & 0.02 & 0.03 \\
Eu II & 0.02 & 0.03 & 0.02 & 0.01 & 0.03 \\
Dy II & 0.02 & 0.03 & 0.01 & 0.03 & 0.04 \\
Er II & 0.02 & 0.03 & $-$0.07 & $-$0.01 & 0.12 \\
Yb II & 0.06 & 0.02 & $-$0.03 & 0.01 & 0.11 \\           
\hline
\end{tabular}%
}
\tablefoot{The full table for all three stars is available online.}
\end{table}

\begin{table}
\caption{Accreted $r_{lim}$ stars based on dynamic selection criteria.}             
\label{table:GSE_Sequoia}      
\centering    
\resizebox{\columnwidth}{!}{%
\begin{tabular}{c c c c}   
\hline\hline       
                      
Stellar ID & GSE member & Sequoia member & Criteria \\ 
\hline                    
   2MASSJ19534978-5940001 & \checkmark & - & \cite{Myeong_2019}, \\ 
   &&& \cite{Feuillet_2021} \\
   2MASSJ19345497-5751400 & \checkmark & - & \cite{Myeong_2019} \\
   2MASSJ20560913-1331176 & \checkmark & - & \cite{Feuillet_2021} \\
   2MASSJ19202070-6627202 & \checkmark & - & \cite{Feuillet_2021} \\
   HD 184266 & \checkmark & - & \cite{Feuillet_2021} \\
   2MASSJ21402305-1227035 & - & \checkmark & \cite{Myeong_2019}, \\
   &&& \cite{Feuillet_2021}\\
   CD-24 1782 & - & \checkmark & \cite{Myeong_2019},\\
   &&& \cite{Feuillet_2021} \\
   
\hline                  
\end{tabular}%
}
\end{table}

\begin{table}
\caption{Lanthanide mass fractions of $r_{lim}$ stars.} \label{table:Xlas}      
\centering    
\begin{tabular}{l r r }   
\hline\hline 
Stellar ID & $X_{La}$ & $\sigma_{X_{La}}$ \\
& & (dex)\\
\hline                    
J21402305-1227035 & $-$2.95 & 0.22 \\
J00385967+2725516 & $-$2.38 & 0.17 \\
J20313531-3127319 & $-$2.17 & 0.15 \\
J10344785-4823544 & $-$2.52 & 0.37 \\
J13085850-2712188 & $-$2.43 & 0.37 \\
J13335283-2623539 & $-$2.51 & 0.37 \\
J05384334-5147228 & $-$2.46 & 0.39 \\
J01094330-5907230 & $-$2.45 & 0.37 \\
J132604.5-152502 & $-$2.50 & 0.15 \\
J160642.3-163245 & $-$2.56 & 0.28 \\
J19594558-2549075 & $-$2.46 & 0.53 \\
J163931.1-052252 & $-$2.65 & 0.32 \\
J14533307-4428301 & $-$2.50 & 0.37 \\
J14435196-2106283 & $-$2.68 & 0.37 \\
J20560913-1331176 & $-$2.54 & 0.53 \\
J18121045-4934495 & $-$2.66 & 0.53 \\
J164551.2-042947 & $-$2.76 & 0.37 \\
J19534978-5940001 & $-$2.61 & 0.53 \\
J035509.3-063711 & $-$2.64 & 0.22 \\
J19202070-6627202 & $-$2.76 & 0.54 \\
J14164084-2422000 & $-$2.81 & 0.54 \\
J19345497-5751400 & $-$2.64 & 0.53 \\
J19494025-5424113 & $-$2.77 & 0.54 \\
J08025449-5224304 & $-$2.93 & 0.40 \\
J03563703-5838281 & $-$2.87 & 0.54 \\
J003052.7-100704 & $-$2.97 & 0.26 \\
J21370807-0927347 & $-$3.00 & 0.54 \\
J17285930-7427532 & $-$3.01 & 0.54 \\
J154755.2-083710 & $-$3.20 & 0.30 \\
CS 22186-023 & $-$2.83 & 0.48 \\
CS 22879-103 & $-$2.29 & 0.30 \\
CS 22891-209 & $-$2.64 & 0.35 \\
CS 22897-008 & $-$3.27 & 0.32 \\
CS 22937-072 & $-$2.47 & 0.36 \\
CS 22940-070 & $-$2.37 & 0.35 \\
CS 22956-114 & $-$2.25 & 0.38 \\
CS 30494-003 & $-$2.30 & 0.34 \\
CD-24 1782   & $-$2.51 & 0.46 \\
G026-001     & $-$2.48 & 0.31 \\
HD 13979     & $-$2.13 & 0.36 \\
HD 19445     & $-$2.43 & 0.48 \\
HD 26169     & $-$2.38 & 0.30 \\
HD 88609     & $-$2.70 & 0.44 \\
HD 122563    & $-$2.71 & 0.37 \\
HD 175606    & $-$2.18 & 0.36 \\
HD 184266    & $-$2.09 & 0.30 \\
HD 237846    & $-$2.62 & 0.42 \\
HE 1320-1339 & $-$2.14 & 0.35 \\
\hline                  
\end{tabular}
\end{table}

\end{appendix}

%
%

\end{document}